\documentclass[aps,pre,10pt,showpacs,floatfix,onecolumn,nofootinbib,a4paper]{revtex4-2}

\usepackage{amssymb, amsmath, amsthm, mathtools}
\usepackage{bm}
\usepackage{mathrsfs}
\usepackage{hyperref,bookmark}
\usepackage{graphicx}
\usepackage{epsfig,color}
\usepackage{float}
\usepackage{subcaption}
\usepackage{verbatim}
\usepackage{dcolumn}
\usepackage{tikz}
\usepackage{url}

\usepackage[utf8x]{inputenc}
\usepackage[english]{babel}
\usepackage{mathrsfs}
\usepackage{epsfig}
\usepackage{listings}

\usepackage{mdwlist}
\usepackage{bm}
\usepackage{oldgerm}
\usepackage{geometry}
\usepackage{relsize}
\usepackage{afterpage}
\usepackage{lmodern}
\usepackage{tabularx}
\usepackage{cancel}
\usepackage[overload]{empheq}
\usepackage{enumitem}

\usepackage{soul}

\newcommand{\dd}{\mathrm{d}}
\newcommand{\ex}{\mathrm{e}}

\newcommand{\vt}{\vartheta}
\newcommand{\vp}{\varphi}
\newcommand{\trans}{^\mathsf{T}}
\newcommand{\transl}{\mathscr{V}}
\newcommand{\av}{\bm{a}}

\newcommand{\cv}{\bm{c}}
\newcommand{\cper}{\bm{c}_\perp}

\newcommand{\dv}{\bm{d}}

\newcommand{\e}{\bm{e}}
\newcommand{\x}{\bm{x}}
\newcommand{\zero}{\bm{0}}
\newcommand{\y}{\bm{y}}

\newcommand{\normal}{\bm{\nu}}
\newcommand{\surface}{\mathscr{S}}
\newcommand{\tplane}{\mathscr{T}}

\newcommand{\nablas}{\nabla\!_\mathrm{s}}
\newcommand{\nablast}{\nabla^\ast\!\!\!_\mathrm{s}}

\newcommand{\nablastwo}{\nabla^2\!\!\!_\mathrm{s}\,}

\newcommand{\uv}{\bm{u}}

\newcommand{\rv}{\bm{r}}
\newcommand{\rvd}{\dot{\rv}}

\newcommand{\bv}{\bm{b}}

\newcommand{\vv}{\bm{v}}

\newcommand{\W}{\mathbf{W}}
\newcommand{\WW}{\mathbb{W}}
\renewcommand{\AA}{\mathbb{A}}

\newcommand{\F}{\mathbf{F}}

\renewcommand{\H}{\mathbf{H}}

\newcommand{\I}{\mathbf{I}}

\newcommand{\R}{\mathbf{R}}
\newcommand{\Rd}{\mathbf{R}_\mathrm{d}}
\newcommand{\Rb}{\mathbf{R}_\mathrm{b}}

\newcommand{\U}{\mathbf{U}}
\newcommand{\V}{\mathbf{V}}
\newcommand{\Id}{\mathbf{I}}

\newcommand{\curl}{\operatorname{curl}}
\newcommand{\curls}{\curl_\mathrm{s}\!}
\newcommand{\divs}{\diver_\mathrm{s}}
\newcommand{\divc}{\diver_\mathrm{c}}
\newcommand{\tr}{\operatorname{tr}}

\newcommand{\skw}{\operatorname{skw}}
\newcommand{\sgn}{\operatorname{sgn}}

\newcommand{\curvature}{(\nablas\normal)}

\newcommand{\A}{\mathbf{A}}

\newcommand{\euclid}{\mathscr{E}}
\newcommand{\framec}{(\e_x,\e_y,\e_z)}

\newcommand{\framen}{(\e_1,\e_2,\normal)}

\newcommand{\frameu}{(\uv_1,\uv_2,\normal)}
\newcommand{\framev}{(\vv_1,\vv_2,\normal^\ast)}

\newcommand{\tangent}{\bm{t}}

\newcommand{\orth}{\mathsf{SO}(3)}

\newcommand{\proj}{\mathbf{P}(\normal)}

\newcommand{\Proj}{\mathbf{P}}

\newcommand{\sphere}{\mathbb{S}^2}

\newcommand{\diver}{\operatorname{div}} 

\newcommand{\high}[1]{{\color{red}{#1}}}
\newcommand{\High}[1]{{\color{magenta}{#1}}}

\newcommand{\ws}{w_\mathrm{s}}
\newcommand{\wdr}{w_\mathrm{d}}
\newcommand{\wb}{w_\mathrm{b}}
\newcommand{\kg}{\kappa_\mathrm{g}}
\newcommand{\spicyl}{\mathscr{P}}
\newcommand{\cyl}{\mathscr{C}}
\newcommand{\cone}{\mathscr{K}}

\setcitestyle{authoryear,round}

\newtheorem{definition}{Definition}

\theoremstyle{definition}
\newtheorem{remark}{Remark}

\begin{document}
	\latintext
	\title{Soft elasticity of soft shells}
	\author{Andr\'e M. Sonnet}
	\email{andre.sonnet@strath.ac.uk}
	\affiliation{Department of Mathematics and Statistics, University of Strathclyde, 26 Richmond Street, Glasgow, G1 1XH, U.K. }
	\author{Epifanio G. Virga}
	\email{eg.virga@unipv.it}
	\affiliation{Department of Mathematics, University of Pavia, Via Ferrata 5, 27100 Pavia, Italy }

	\date{\today}

	\begin{abstract}
\emph{Soft} modes of a shell are one-parameter families of non-trivial deformations that preserve the elastic energy. We identify all soft modes of the \emph{soft shells} described by the variational theory of  put forward by \citet{sonnet:variational}. They are seen to be strictly connected with the \emph{Bonnet families}, a special class of solutions to the problem posited by Bonnet, who asked whether there exist non-trivial isometries that preserve the mean curvature. Differential geometry deems as equivalent all surfaces in one and the same Bonnet family, but continuum mechanics is interested more specifically in soft deformation modes  and the shapes they traverse, as they may describe ways to ignite  fluctuations of soft shells that ideally require no energy to be activated. We give a general criterion for the existence of soft modes and a method to construct them based on a cascade of path integrations. We apply both criterion and method to some exemplary cases.
	\end{abstract}

	\maketitle

\section{Introduction}\label{sec:intro}
\emph{Softness} is not a quality that one would naturally associate with ordinary shells. Here, a shell is considered to be as a  \emph{material surface} capable of having its constituting fibers easily twisted in their local tangent plane. As such, our definition of softness pertains more to kinematics than to constitutive stipulations.

The deformation mode that qualifies a soft shell is called \emph{drilling} \citep{hughes:drilling} and is associated with averaged relative rotations about the shell's normal, while \emph{bending} is the mode associated with averaged relative rotations about an axis in the shell's tangent plane \citep{sonnet:variational}. These modes supplement the standard \emph{stretching} mode, contemplated in classical membrane theory, that is only related to the \emph{metric} changes induced by the deformation.

In \citet{sonnet:variational}, we proposed a strain elastic energy for soft shells to which each individual deformation mode brings its own contribution,
\begin{equation}
\label{eq:energy_density}
w=\tfrac12\mu_\mathrm{s}\ws+\tfrac12\mu_\mathrm{d}\wdr+\tfrac14\mu_\mathrm{b}\wb,
\end{equation}
where $\mu_\mathrm{s}$, $\mu_\mathrm{d}$, and $\mu_\mathrm{b}$ are (positive) elastic moduli weighing the stretching, drilling, and bending energies $\ws$, $\wdr$, and $\wb$, respectively (see Sect.~\ref{sec:kino}).\footnote{The factor $\frac14$ is in \eqref{eq:energy_density} just to remind us that $\wb$ is \emph{quartic} in the appropriate deformation measure, while the remaining energies are quadratic.} 

We say that a deformation is \emph{stretching}, \emph{drilling}, or \emph{bending neutral}, whenever $\ws$, $\wdr$, or $\wb$ vanishes, respectively. Globally neutral deformations, for which  $w$ vanishes identically, are instances of \emph{soft elasticity}, where the shell's shape can change with no elastic cost. A \emph{single} deformation with this property will be called \emph{soft elastic}. More specifically, a \emph{soft mode} will designate a one-parameter family of globally neutral deformations.

 Given that all three different deformation modes are kinematically independent, one might think that they are also energetically independent. This, however, is not the case. There are indeed special deformations for which if $\ws$ vanishes then also both $\wdr$ and $\wb$ do, thus delivering an instance of soft elasticity, as is the case for conformal deformations of a minimal surface onto another \citep{sonnet:neutral,tiwari:orthogonal}. Moreover, as already anticipated in the conclusions of \citet{sonnet:neutral}, stretching neutral modes are drilling neutral not only for deformations of minimal surfaces but for deformations of  \emph{all} surfaces;  proofs are given by \citet{sonnet:metric} and \citet{tiwari:orthogonal}.\footnote{The proof of \citet{sonnet:metric} is based on Cartan's moving frame formalism, while that of \citet{tiwari:orthogonal} is obtained by decomposing an invariant third-rank tensor $\H$, which features in both $\wdr$ and $\wb$, into two orthogonal components, one tangential and the other normal to the surface representing the shell. The tangential component  contributes only to $\wdr$, whereas the normal component  contributes only to $\wb$.}

This paper is concerned with soft modes and the reference configurations that admit them, according to the theory of \citet{sonnet:variational}. The admissible reference configurations are precisely the \emph{Bonnet surfaces}; they solve the problem posited in 1867 by Bonnet about the existence of surfaces endowed with non-trivial isometries preserving the mean curvature. The Bonnet problem is still partly open,\footnote{A very recent advancement for \emph{compact} surfaces can be found in \citep{bobenko:compact}.} Here, we are rather more interested in the explicit construction of the soft modes of a Bonnet surface.

The paper is organized as follows. In Sect.~\ref{sec:kino}, we recall  some basic kinematic prerequisites to make the paper self-contained. In Sect.~\ref{sec:drill}, we prove a novel formula for the drilling energy $\wdr$ that renders its intrinsic nature immediately obvious. In Sect.~\ref{sec:soft}, we characterize soft elastic deformations in general and present a method based on a cascade of path integrations that enable us to construct one, whenever it exists. Section~\ref{sec:bonnet} is about Bonnet surfaces and their properties relevant to our study. Here we give our criterion for the existence of soft modes and apply it to specific Bonnet surfaces. In Sect.~\ref{sec:applications}, to provide concrete applications, we construct the soft modes of selected Bonnet surfaces. 
Finally, in Sect.~\ref{sec:conclusion}, we collect the conclusions of this work and comment briefly on  possible ways to extend it. The paper is closed by two technical  Appendices containing ancillary calculations.

\section{Kinematics}\label{sec:kino}
In this section, we recall from \citep{sonnet:bending-neutral,sonnet:metric} the properties of surface deformations strictly needed here; the reader is referred to the original papers for both definitions and basic mathematical tools of differential  calculus on surfaces.

We identify the reference configuration of a thin shell with a \emph{smooth} surface $\surface$ (at least of class $C^4$) in three-dimensional space $\euclid$, whose translation space will be denoted by $\transl$. $\surface$ is assumed to be orientable and its orientation is described by the unit normal field $\normal$.

\begin{figure}[t] 
	\centering
		\includegraphics[width=.75\linewidth]{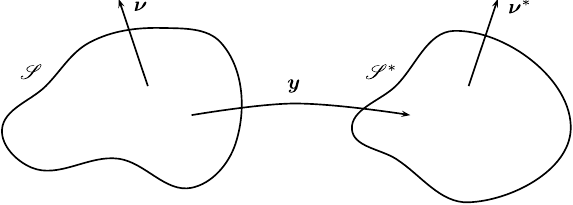}
	\caption{The deformation $\y$ maps $\surface$ onto $\surface^\ast$; the corresponding unit normal fields are denoted $\normal$ and $\normal^\ast$, respectively.}
	\label{fig:diagram}
\end{figure}

A deformation is a \emph{smooth} mapping  $\y:\surface\to\euclid$ (at least continuously twice-differentiable), whose image $\surface^\ast:=\y(\surface)$ is the present configuration of the shell. $\surface^\ast$ is a surface (at least of class $C^2$) oriented by the unit normal $\normal^\ast$, as shown in Fig.~\ref{fig:diagram}.

The deformation (surface) gradient $\nablas\y$ can uniquely be decomposed as
\begin{equation}
	\label{eq:deformation_gradient}
	\nablas\y=\R\U=\V\R,
\end{equation}	
where $\U$ and $\V$ are the \emph{stretching} tensors and $\R\in\orth$ is a  rotation \citep{man:coordinate}. When restricted to the plane $\tplane_{\x}$ tangent to $\surface$ at the point $\x$, $\U$ is a positive definite symmetric second-rank tensor, as is $\V$ restricted to the plane $\tplane^\ast_{\y(\x)}$ tangent to $\surface^\ast$ at $\y(\x)$; they have the same eigenvalues (principal stretches) $\lambda_1$ and $\lambda_2$, and eigenframes $(\uv_1,\uv_2)$, $(\vv_1,\vv_2)$, respectively. Accordingly, $\nablas\y(\x)$ restricted to $\tplane_{\x}$ is a linear mapping from $\tplane_{\x}$ onto $\tplane^\ast_{\y(\x)}$.

Here, as in \citep{sonnet:metric}, we regard all tensors in \eqref{eq:deformation_gradient} as \emph{extended} to linear mappings of $\transl$ into itself, with  $\nablas\y$ vanishing on $\normal$ and both $\U$ and $\V$ symmetric and vanishing on $\normal$ and $\normal^\ast$, respectively. Thus, the following representations apply,
\begin{subequations}
	\begin{align}
		\U&=\lambda_1\uv_1\otimes\uv_1+\lambda_2\uv_2\otimes\uv_2,\label{eq:U}\\
		\V&=\lambda_1\vv_1\otimes\vv_1+\lambda_2\vv_2\otimes\vv_2,\label{eq:V}\\
		\R&=\vv_1\otimes\uv_1+\vv_2\otimes\uv_2+\normal^\ast\otimes\normal,\label{eq:R}
	\end{align}
\end{subequations}
and we stipulate that $\normal=\uv_1\times\uv_2$ and that the moving frames $\frameu$ and $\framev$, on $\surface$ and $\surface^\ast$ respectively, are equally oriented, so that $\normal^\ast=\vv_1\times\vv_2$.

Equation \eqref{eq:deformation_gradient} is the \emph{polar} decomposition of the deformation gradient and $\R$ is thus called the \emph{polar} rotation of $\nablas\y$. It was shown by \citet{sonnet:bending-neutral} how $\R$ can be \emph{uniquely} decomposed in the product
\begin{equation}
	\label{eq:R_decomposition}
	\R=\Rb\Rd,
\end{equation}
where $\Rd$, the \emph{drilling} rotation, has axis along the normal $\normal$, whereas $\Rb$, the \emph{bending} rotation, has axis on the local tangent plane to $\surface$.

Drilling and bending energies are derived by \citet{sonnet:variational} by taking local relative averages of the square moduli $d^2$ and $b^2$ of the vectors $\dv$ and $\bv$ representing $\Rd$ and $\Rb$, respectively, in \citeauthor{rodrigues:lois}' (\citeyear{rodrigues:lois}) vectorial representation of rotations \citep{sonnet:metric}.

The invariant measures of drilling and bending thus obtained are
\begin{align}
	\wdr&=|\W(\normal)\circ\H|^2\label{eq:w_d},\\
	\wb&=\left(|\H|^2-\frac12|\W(\normal)\circ\H|^2-4\normal\cdot\H\circ\nablas\normal\right)^2,\label{eq:w_b}
\end{align}
where $\H$ is the \emph{invariant rotation gradient}, a third-rank tensor given by
\begin{equation}
	\label{eq:H_definition}
	\H:=\R\trans\nablas\R,
\end{equation}
$\nablas\normal$ is the curvature tensor of $\surface$, and the \emph{left} and \emph{right} circle contractions $\circ$ to a vector  in \eqref{eq:w_d} and \eqref{eq:w_b}, respectively, are defined as
\begin{equation}
\label{eq:circles_definition}
\A\circ(\av\otimes\bv\otimes\cv):=(\av\cdot\A\bv)\cv\quad\text{and}\quad(\av\otimes\bv\otimes\cv)\circ\A:=(\bv\cdot\A\cv)\av,
\end{equation}
for any second-rank tensor $\A$ and any third-rank triad $\av\otimes\bv\otimes\cv$ \citep{sonnet:variational}.

Here and in the following, for any vector $\vv\in\transl$, $\W(\vv)$ denotes the second-rank skew-symmetric tensor such that $\W(\vv)\uv=\vv\times\uv$, for all $\uv\in\transl$. Specifically, for $\vv=\normal$, $W(\normal)$ can equivalently be written as
\begin{equation}\label{eq:W_representation}
\W(\normal)=\uv_2\otimes\uv_1-\uv_1\otimes\uv_2,
\end{equation}	
while the projection onto the local tangent plane of $\surface$ is  $\proj:=\I-\normal\otimes\normal$, where $\I$ is the identity acting on $\transl$.

The \emph{stretching} energy density $\ws$ can be given either the standard form
\begin{equation}
	\label{eq:w_s}
	\ws=|\U-\proj|^2=|\V-\mathbf{P}(\normal^\ast)|^2,
\end{equation}
or any other that would imply that $\ws\equiv0$ on \emph{isometric} deformations, for which $\U\equiv\proj$ and $\V\equiv\Proj(\normal^\ast)$.

\citet{sonnet:metric} proved that in the moving frame $\frameu$ the invariant rotation gradient $\H$ can be represented in the form
\begin{equation}
	\label{eq:H_representation}
	\H=\W(\uv_1)\otimes\av_1+\W(\uv_2)\otimes\av_2+\W(\normal)\otimes\av_3,
\end{equation}
where $\av_1$, $\av_2$, and $\av_3$ are vector fields tangent to $\surface$.

The moving frame $\frameu$ glides over $\surface$ subject to the (structure) laws
\begin{equation}\label{eq:structure_laws}
	\begin{cases}
		\nablas\uv_1=\uv_2\otimes\cv+\normal\otimes\dv_1,\\
		\nablas\uv_2=-\uv_1\otimes\cv+\normal\otimes\dv_2,\\
		\nablas\normal=-\uv_1\otimes\dv_1-\uv_2\otimes\dv_2,
	\end{cases}
\end{equation}
where the \emph{tangential}\footnote{We say that a vector field $\vv$ is \emph{tangential} whenever $\vv\cdot\normal\equiv0$.} fields $\cv$, $\dv_1$, and $\dv_2$ are the \emph{connectors} of the frame; $\cv$ is the \emph{spin} connector, while $\dv_1$ and $\dv_2$ are the \emph{curvature} connectors. Since $\nablas\normal$ is symmetric, the latter connectors must satisfy the identity
\begin{equation}
	\label{eq:connector_identity}
	\dv_1\cdot\uv_2=\dv_2\cdot\uv_1.
\end{equation}
Moreover, the third equation in \eqref{eq:structure_laws} implies that
\begin{subequations}\label{eq:mean_and_gaussian}
	\begin{align}
		2H:=&\tr\curvature=-(\dv_1\cdot\uv_1+\dv_2\cdot\uv_2)\label{eq:mean_curvature},\\
		K:=&\det\curvature=\dv_1\times\dv_2\cdot\normal,\label{eq:gaussian_curvature}
	\end{align}
\end{subequations}
where $H$ and $K$ are the \emph{mean} and \emph{Gaussian} curvatures of $\surface$, respectively.
\begin{remark}
	\label{rmk:colares}
Following \citet{colares:isometric}, by \eqref{eq:connector_identity}, we can represent the curvature connectors $(\dv_1,\dv_2)$ in terms of $H$, $K$, and an angle $\chi$. It follows from \eqref{eq:connector_identity} and \eqref{eq:mean_curvature} that 
\begin{subequations}
\label{eq:d_representations}
	\begin{align}[left={\empheqlbrace}]
	&\dv_1=(\xi-H)\uv_1+\eta\uv_2,\label{eq:d_1_representation}\\
	&\dv_2=\eta\uv_1-(\xi+H)\uv_2,\label{eq:d_2_representation}
	\end{align}
\end{subequations}
where $\xi$ and $\eta$ are parameters to be determined. By inserting \eqref{eq:d_representations} in \eqref{eq:gaussian_curvature}, we see that
\begin{equation}
\label{eq:xi_eta_circle}
\xi^2+\eta^2=H^2-K\geqq0,
\end{equation}
and so we can represent $\xi$ and $\eta$ as
\begin{equation}
\label{eq:xi_eta}
\xi=\sqrt{H^2-K}\cos\chi,\qquad\eta=\sqrt{H^2-K}\sin\chi.
\end{equation}
Inserting \eqref{eq:xi_eta} into \eqref{eq:d_representations} delivers the desired representation.
\end{remark}
\begin{remark}
		\label{rmk:integrability}
	As recalled by \citet{sonnet:bending-neutral}, for a second-rank tensor field $\F$ defined on $\surface$ so that $\F\normal\equiv\zero$, there exists a vector field $\vv$ on $\surface$ such that $\F=\nablas\vv$, if and only if\footnote{Here skw acts as follows on a generic \emph{triadic} tensor $\vv_1\otimes\vv_2\otimes\vv_3$,
		\begin{equation*}
			\label{eq:skw_third_rank_tensor}
			2\skw(\vv_1\otimes\vv_2\otimes\vv_3):=\vv_1\otimes\vv_2\otimes\vv_3-\vv_1\otimes\vv_3\otimes\vv_2.
		\end{equation*}
		The (easier) vectorial form of \eqref{eq:integrability_vector} is also recorded in \eqref{eq:integrability_scalar}, where $\skw$ acts in the standard way on any \emph{dyadic} tensor $\vv_1\otimes\vv_2$.}
	\begin{equation}
		\label{eq:integrability_vector}
		\skw(\nablas\F)=\skw(\F\curvature\otimes\normal).
	\end{equation}
	By identifying $\F$ in turn with $\nablas\uv_1$, $\nablas\uv_2$, and $\nablas\normal$ in \eqref{eq:structure_laws}, setting 
	\begin{equation}
		\label{eq:c_d_1_components}
		\cv=c_1\e_1+c_2\e_2,\ \dv_1=d_{11}\e_1+d_{12}\e_2,\ \dv_2=d_{21}\e_1+d_{22}\e_2,\quad \text{with}\quad  d_{12}=d_{21},
	\end{equation}
	\citet{sonnet:metric} derived from \eqref{eq:integrability_vector} the following (necessary and sufficient) integrability conditions,
	\begin{equation}
		\label{eq:integrability_identities}
		\begin{cases}
			\curls\cv+\dv_1\times\dv_2=c_1\dv_1\times\normal+c_2\dv_2\times\normal=\normal\times\curvature\cv,\\
			\curls\dv_1-\cv\times\dv_2=d_{11}\dv_1\times\normal+d_{12}\dv_2\times\normal=\normal\times\curvature\dv_1,\\
			\curls\dv_2+\cv\times\dv_1=d_{21}\dv_1\times\normal+d_{22}\dv_2\times\normal=\normal\times\curvature\dv_2,
		\end{cases}
	\end{equation}
	where the last equalities follow easily from \eqref{eq:structure_laws} and \eqref{eq:c_d_1_components}.  
	\end{remark}
	\begin{remark}
		\label{rmk:simplified_a_equations}
		For any smooth \emph{tangential} vector field $\av$, the surface component of $\curls\av$ is only determined by $\av$ and the curvature tensor,
		\begin{equation}
			\label{eq:curl_tangential_component}
			\proj\curls\av=\normal\times\curvature\av.
		\end{equation}
		The proof of \eqref{eq:curl_tangential_component} is a simple exercise; it can be obtained by decomposing $\av$ in the frame $\frameu$ as $\av=a_1\uv_1+a_2\uv_2$ to show then with the aid of \eqref{eq:structure_laws} that
		\begin{equation}
			\label{eq:curl_tangential_components}
			\proj\curls\av=(a_1d_{12}+a_2d_{22})\uv_1-(a_1d_{11}+a_2d_{21})\uv_2,
		\end{equation}
		which by \eqref{eq:c_d_1_components} is equivalent to \eqref{eq:curl_tangential_component}.
	\end{remark}
	\begin{remark}
		\label{rmk:gauss_mainardi}
		In view of \eqref{eq:curl_tangential_component}, equations \eqref{eq:integrability_identities} are thus equivalent to
		\begin{subequations}\label{eq:gauss_mainardi}
			\begin{align}[left={\empheqlbrace}]
				&\curls\cv\cdot\normal=-K,\label{eq:gauss_equation}\\
				&\curls\dv_1\cdot\normal=\cper\cdot\dv_2,\label{eq:mainardi_equation_1}\\
				&\curls\dv_2\cdot\normal=-\cper\cdot\dv_1,	\label{eq:mainardi_equation_2}
			\end{align}
		\end{subequations}
		where
		\begin{equation}
			\label{eq:c_perp_definition}
			\cper:=\normal\times\cv
		\end{equation}
		and \eqref{eq:gaussian_curvature} has also been used. Equations \eqref{eq:gauss_mainardi} are fundamental in the differential geometry of surfaces in three-dimensional space: \eqref{eq:gauss_equation} is the Gauss equation, while \eqref{eq:mainardi_equation_1} and \eqref{eq:mainardi_equation_2} are the Peterson-Mainardi-Codazzi equations \citep[p.\,448]{needham:visual}.
	\end{remark}
	\begin{remark}
		\label{rmk:c=0}
		It is an immediate consequence of \eqref{eq:gauss_equation} that $\cv\equiv\zero$ implies $K\equiv0$.
\end{remark}
By use of \eqref{eq:structure_laws} and \eqref{eq:connector_identity}, \citet{sonnet:metric} gave $\wdr$ and $\wb$ in \eqref{eq:w_d} and \eqref{eq:w_b}
the following expressions in terms of the vectors $\av$,
\begin{equation}
	\label{eq:wd_wb_a}
	\wdr=4a_3^2\quad\text{and}\quad\wb=4\left(a_1^2+a_2^2+2(\av_1\cdot\dv_2-\av_2\cdot\dv_1) \right)^2,
\end{equation}
which show that $\wdr$ bears \emph{two} degrees of freedom (the scalar components of $\av_3$), while $\wb$ bears \emph{four} degrees of freedom (the components of $\av_1$ and $\av_2$). 

\citet{sonnet:metric}  also proved (see, in particular Appendix~D) that the vectors $\av$ must satisfy the following scalar equations,
\begin{subequations}\label{eq:integrability_conditions}
	\begin{align}[left={\empheqlbrace}]
		&\lambda_1(\dv_1-\av_2)\cdot\uv_2=\lambda_2(\dv_2+\av_1)\cdot\uv_1, \label{eq:integrability_condition_1}\\
		&\lambda_1(\av_3\cdot\uv_2)+(\lambda_1-\lambda_2)(\cv\cdot\uv_2)-\nablas\lambda_2\cdot\uv_1=0, \label{eq:integrability_condition_2}\\
		&\lambda_2(\av_3\cdot\uv_1)+(\lambda_2-\lambda_1)(\cv\cdot\uv_1)+\nablas\lambda_1\cdot\uv_2=0.\label{eq:integrability_condition_3}
	\end{align}
\end{subequations}
These follow, also with the aid of \eqref{eq:structure_laws}, from requiring that $\R\U$, with $\U$ as in \eqref{eq:U} and $\R$ as in \eqref{eq:R}, is indeed a surface gradient.

For an isometric deformation $\lambda_1=\lambda_2=1$, and so equations \eqref{eq:integrability_conditions} and \eqref{eq:connector_identity} imply that 
\begin{subequations}\label{eq:isometry_implications}
\begin{align}[left={\empheqlbrace}]
	&\av_3=\zero,\label{eq:isometry_implications_a}\\
	&\av_1\cdot\uv_1+\av_2\cdot\uv_2=0.\label{eq:isometry_implications_b}
\end{align}
\end{subequations}
By \eqref{eq:w_s} and \eqref{eq:wd_wb_a}, equation \eqref{eq:isometry_implications_a}  in particular shows that
\begin{equation}
	\ws=0\quad\Rightarrow\quad\wdr=0.
\end{equation}

Different moving frames have different connectors. In particular,  if we let $\framen$ be  the eigenframe of the curvature tensor $\nablas\normal$, so that
\begin{equation}
	\label{eq:curvature_diagonal}
	\nablas\normal=\kappa_1\e_1\otimes\e_1+\kappa_2\e_2\otimes\e_2,
\end{equation}
where $\kappa_1$ and $\kappa_2$ are the principal curvatures of $\surface$, by comparing \eqref{eq:curvature_diagonal} and \eqref{eq:structure_laws}, we see that in this frame
\begin{equation}
	\label{eq:d_connectors_diagonal}
	\dv_1=-\kappa_1\e_1\quad\text{and}\quad\dv_2=-\kappa_2\e_2,
\end{equation}
so that equations \eqref{eq:structure_laws} are replaced by
	\begin{equation}
	\label{eq:structure_laws_curvature}
	\begin{cases}
		\nablas\e_1=\e_2\otimes\cv-\kappa_1\normal\otimes\e_1,\\
		\nablas\e_2=-\e_1\otimes\cv-\kappa_2\normal\otimes\e_2,\\
		\nablas\normal=\kappa_1\e_1\otimes\e_1+\kappa_2\e_2\otimes\e_2.
	\end{cases}
\end{equation}
Moreover, for the frame $\framen$ equations \eqref{eq:mainardi_equation_1} and \eqref{eq:mainardi_equation_2} become 
\begin{subequations}\label{eq:kappa_c_equations}
\begin{align}[left={\empheqlbrace}]
&\e_2\cdot\nablas\kappa_1=(\kappa_1-\kappa_2)(\e_1\cdot\cv),\label{eq:kappa_c_equations_2}\\
&\e_1\cdot\nablas\kappa_2=(\kappa_1-\kappa_2)(\e_2\cdot\cv).\label{eq:kappa_c_equations_3}
\end{align}
\end{subequations}

Finally, if $\y$ is an isometric deformation of $\surface$, the deformed surface $\surface^\ast$ has Gaussian curvature $K^\ast=K$, by Gauss' \emph{theorema egregium} \citep[p.\,140]{needham:visual}, and mean curvature $H^\ast$ given by \citep{sonnet:metric}
\begin{equation}
	\label{eq:H_star}
	2H^\ast=2H+\av_2\cdot\e_1-\av_1\cdot\e_2,
\end{equation}
where the $\av$ vectors are those associated with the frame $\framen$ via the appropriate form of \eqref{eq:H_representation}.\footnote{Accordingly, for an isometry, \eqref{eq:isometry_implications_b} requires that $\av_1\cdot\e_1+\av_2\cdot\e_1=0$.}

Equations \eqref{eq:integrability_condition_2} and \eqref{eq:integrability_condition_3} will be instrumental in the following section to find  for $\wdr$ an intrinsic formula simpler than \eqref{eq:w_d}. A formula for $\wb$ simpler than \eqref{eq:w_b} had already been obtained in \citet{sonnet:variational},
\begin{equation}
	\label{eq:w_b_curvature}
	\wb=4\left[|\V(\nablast\normal^\ast)|^2-|\nablas\normal|^2\right]^2,
\end{equation}
where $\nablast\normal^\ast$ is the curvature tensor of $\surface^\ast$. For an isometric deformation, since $K^\ast=K$ and $\V=\Proj(\normal^\ast)$, $\wb$ acquires the even simpler form
\begin{equation}
	\label{eq:w_b_isometry}
	\wb=4(4H^{\ast2}-4H^2)^2.
\end{equation}
 
\section{Intrinsic Drilling Energy}\label{sec:drill}
It was neatly shown by \citet{tiwari:orthogonal} that $\wdr$ is uniquely determined by the stretching tensor $\U$ and its surface gradient $\nablas\U$, this latter third-rank tensor entering explicitly their formula (116). Here, we  express $\wdr$ in an equivalent, clearly intrinsic form, which does not involve explicitly the third-rank tensor field $\nablas\U$, but rather the \emph{covariant} divergence of a second-rank tensor field constructed from $\U$. 

Translated in the language of this paper, the result proved by \citet{tiwari:orthogonal} amounts to write $\av_3$ in the form
\begin{equation}
	\label{eq:a_3_fried}
	\av_3=\frac{1}{\det\U}(\U\nablas\U)\circ\W(\normal),
\end{equation} 
valid for any deformation $\y$ (see Appendix~\ref{sec:alternative} for an independent proof of this equation). Inserting \eqref{eq:a_3_fried} into the first equation in \eqref{eq:wd_wb_a} delivers equation (116) of \citet{tiwari:orthogonal}.

Here we give a simpler, equivalent formula for $\av_3$, which has the net advantage of exhibiting the \emph{intrinsic} nature of this vector field (and consequently of $\wdr$). We start by remarking that
\begin{equation}
	\label{eq:step_1}
	\U\W(\normal)=-\lambda_1\uv_1\otimes\uv_2+\lambda_2\uv_2\otimes\uv_1,
\end{equation}
which follows from \eqref{eq:U} and \eqref{eq:W_representation}. By direct computation, we then obtain from \eqref{eq:step_1} that 
\begin{equation}
		\label{eq:step_2}
\begin{split}
\divs(\U\W(\normal))=&-[\nablas\lambda_1\cdot\uv_2+(\lambda_2-\lambda_1)\cv\cdot\uv_1]\uv_1\\
&+[\nablas\lambda_2\cdot\uv_1+(\lambda_2-\lambda_1)\cv\cdot\uv_2]\uv_2+(\lambda_2\dv_2\cdot\uv_1-\lambda_1\dv_1\cdot\uv_2)\normal,
\end{split}
\end{equation}
where we have also made use of \eqref{eq:structure_laws}. By inserting \eqref{eq:integrability_condition_2} and \eqref{eq:integrability_condition_3} in \eqref{eq:step_2}, we easily arrive at 
\begin{equation}
	\label{eq:step_3}
\proj\divs(\U\W(\normal))=\lambda_2(\av_3\cdot\uv_1)\uv_1+\lambda_1(\av_3\cdot\uv_2)\uv_2,
\end{equation}
which by \eqref{eq:U} delivers
\begin{equation}
	\label{eq:a_3_simple}
	\av_3=\frac{1}{\det\U}\U\proj\divs(\U\W(\normal)),
\end{equation}
since $\det\U=\lambda_1\lambda_2$.

Equation \eqref{eq:a_3_simple} is the desired simpler form of $\av_3$, which by \eqref{eq:wd_wb_a} gives an expression for $\wdr$ easier to compute, as it does not require the full third-rank tensor $\nablas\U$ featuring in \eqref{eq:a_3_fried}.	

\begin{remark}
\label{rmk:a_3_covariant}
Since, as discussed for example by \citet[p.\,243]{needham:visual}, the projection onto the local tangent plane of the surface divergence operator $\divs$ delivers the \emph{covariant} divergence $\divc$, equation \eqref{eq:a_3_simple} also reads as
\begin{equation}
	\label{eq:a_3_covariant}
	\av_3=\frac{1}{\det\U}\U\divc(\U\W(\normal)).
\end{equation}
Covariant differentiation is defined in terms of Levi-Civita's parallel transport and it is thus intrinsic \citep[\S23.2]{needham:visual}, which makes $\wdr$ intrinsic too, and so $\wdr$ is related only to metric properties of the deformation.
\end{remark}
\begin{remark}
	\label{rmk:a_3_alternative}
Yet another way of writing $\av_3$ follows from the identity
\begin{equation}
\label{eq:curl_U_identiy}
\divs(\U\W(\normal))=(\curls\U)\trans\normal,
\end{equation}
which shows that only $\curls\U$ actually features in \eqref{eq:a_3_fried}. In particular, since $\U$ is positive definite, equations \eqref{eq:a_3_simple} and \eqref{eq:curl_U_identiy} imply that $\av_3=\zero$ if and only if $\normal\times(\curls\U)\trans\normal=\zero$.
\end{remark}

\section{Soft Elastic Deformations}\label{sec:soft}
We are interested in \emph{deformations} of shells, not just in their changes of \emph{shape}. This is perhaps what most starkly distinguishes kinematics of shells from differential geometry of surfaces: we want to know precisely where individual body-points constituting a material surface are mapped by a deformation, not just in what shape they are arranged. Were we only interested in the latter, we could take advantage of the classical methods of differential geometry of surfaces in three-dimensional space that tailor coordinates to a selected need. We shall not have such liberty: once coordinates are set in $\surface$, we need to follow the journey of all its body-points onto $\surface^\ast$. This will make our kinematic tasks possibly more difficult compared to their mere geometric counterparts.

In the elastic theory of soft shells, there is one issue that deserves special attention, that is, whether or not the theory allows for \emph{soft elasticity}, by which we specifically mean the existence of a full continuum of configurations all sharing the same stored strain energy. Deformations connecting continuously configurations in such a crowded ground state would expend no energy and the shell could freely fluctuate from one configuration to another in that class, if not hampered by external agents. 

Explicit examples of soft elasticity are known from the theory of nematic elastomers; they are accurately recounted in the book by \citet[$\S$\,7.2]{warner:liquid}.\footnote{The reader is also referred to the pioneering works of \citet{golubovic:nonlinear} and \citet{olmsted:rotational}.}

In this section, we shall provide a method to identify both surfaces capable of soft elasticity within our theory of soft shells and, more importantly, also the deformations revealing their free (unenergetic) fluctuations in space.

We say that a deformation is \emph{stretching neutral} if $\ws=0$, \emph{drilling neutral} if $\wdr=0$, and \emph{bending neutral} if $\wb=0$. Thus, in our language, soft elasticity is characterized by the existence of surfaces that may undergo deformations that are at once stretching, drilling, and bending neutral.

As already remarked, stretching neutral deformations are isometric and these are drilling neutral too, so that soft elasticity requires the existence of surfaces for which there are non-trivial\footnote{That is, different from  uniform rotations or translations in space.} bending neutral isometric deformations, each of which we shall call \emph{soft elastic}. By \eqref{eq:w_b_isometry}, this amounts to require the deformation to preserve the mean curvature of $\surface$, so that $H^\ast=H$.

Not all surfaces admit such deformations; those that do are called the \emph{Bonnet surfaces}, as \citet{bonnet:memoire_II} first studied them. He proved that analytic surfaces with no umbilics, that is, points with equal principal curvatures, and \emph{constant} mean curvature are Bonnet surfaces. 

More about Bonnet surfaces will be said in Sect.~\ref{sec:bonnet} below. Here, we shall present a general method for constructing the soft elastic deformations admitted by a Bonnet surface.
\begin{remark}
	\label{rmk:minimal_surfaces}
	Before progressing with this task, we note that minimal surfaces (for which $H\equiv0$) are special cases of  Bonnet surfaces; their soft elastic deformations are embodied by \emph{Bonnet transformations}.\footnote{In the theory of minimal surfaces, a Bonnet transformation is responsible, in particular, for changing a catenoid into a helicoid, see, for example, pp.\,102 and 149 of the book by \citet{dierkes:minimal} or pp.\,237--239 of the book by \cite{jensen:surfaces}.} As shown in \citep{sonnet:metric}, Bonnet transformations of minimal surfaces are the only admissible soft  \emph{pure drilling} deformations, that is, deformations for which there is a frame where we can set $\Rb=\Id$ in \eqref{eq:R_decomposition}. If soft elastic deformations are so constrained, then minimal surfaces are the only admissible Bonnet surfaces.
\end{remark}
In the moving frame $\framen$ of principal directions of curvature, by \eqref{eq:isometry_implications} and \eqref{eq:H_star}, a soft elastic deformation is characterized by having
\begin{subequations}
	\label{eq:H_preserving_isometry}
		\begin{align}[left={\empheqlbrace}]
		&\av_3=\zero, \label{eq:H_preserving_isometry_1}\\
		&\av_1\cdot\e_1+\av_2\cdot\e_2=0, \label{eq:H_preserving_isometry_2}\\
		&\av_1\cdot\e_2-\av_2\cdot\e_1=0.\label{eq:H_preserving_isometry_3}
	\end{align}
\end{subequations}
This frame is governed by \eqref{eq:structure_laws_curvature} and, by \eqref{eq:H_representation}, for an isometry, the invariant rotation gradient $\H$ is represented as
\begin{equation}
	\label{eq:H_isometry_curvature}
	\H=\W(\e_1)\otimes\av_1+\W(\e_2)\otimes\av_2,
\end{equation}
where \eqref{eq:H_preserving_isometry_1} has also been used. 

We shall denote $\vv_i=\R\e_i$, keeping in mind that these unit vectors tangent to $\surface^\ast$ need not coincide with the principal directions of curvature $\e_i^\ast$ of $\surface^\ast$; by \eqref{eq:R}, $\R$ is given by
\begin{equation}
	\label{eq:R_curvature_frame}
	\R=\vv_1\otimes\e_1+\vv_2\otimes\e_2+\normal^\ast\otimes\normal.
\end{equation}
Thus \eqref{eq:H_definition} delivers
\begin{equation}
	\label{eq:nabla_R}
	\nablas\R=\vv_1\otimes\normal\otimes\av_2-\vv_2\otimes\normal\otimes\av_1
	-\normal^\ast\otimes\e_1\otimes\av_2+\normal^\ast\otimes\e_2\otimes\av_1.
\end{equation}

With the aid of \eqref{eq:nabla_R}, \eqref{eq:R_curvature_frame}, \eqref{eq:structure_laws_curvature}, we can express
the surface gradients of the vectors in the frame $\framev$ regarded as fields on $\surface$,
\begin{equation}
	\label{eq:nabla_v}
	\begin{cases}
		\nablas\vv_1=\vv_2\otimes\cv-\kappa_1\normal^\ast\otimes\e_1-\normal^\ast\otimes\av_2,\\
		\nablas\vv_2=-\vv_1\otimes\cv-\kappa_2\normal^\ast\otimes\e_2+\normal^\ast\otimes\av_1,\\
		\nablas\normal^\ast=\kappa_1\vv_1\otimes\e_1+\kappa_2\vv_2\otimes\e_2+\vv_1\otimes\av_2-\vv_2\otimes\av_1.
	\end{cases}
\end{equation}

By applying, again  with the aid of \eqref{eq:kappa_c_equations}, the integrability condition \eqref{eq:integrability_vector} to the tensor fields in \eqref{eq:nabla_v}, by \eqref{eq:curl_tangential_component}, we conclude that the latter are integrable if and only if
\begin{subequations}
	\label{eq:integrability_conditions_rotation}
		\begin{align}[left={\empheqlbrace}]
		&\kappa_1\e_1\times\av_1+\kappa_2\e_2\times\av_2+\av_2\times\av_1=\zero, \label{eq:integrability_conditions_rotation_1}\\
		&\curls\av_1\cdot\normal=\cv_\perp\cdot\av_2, \label{eq:integrability_conditions_rotation_bis_2}\\
		&\curls\av_2\cdot\normal=-\cv_\perp\cdot\av_1,\label{eq:integrability_conditions_rotation_bis_3}
	\end{align}
\end{subequations}
where $\cv$ is the spin connector that features in \eqref{eq:structure_laws_curvature}.

\begin{remark}
	\label{rmk:reassurance}
	It is reassuring to known that with a little more labour one can prove that \eqref{eq:integrability_conditions_rotation} are also the necessary and sufficient conditions for the validity of the integrability identity
	\begin{equation}
		\label{eq:four_rank_integrability}
		2\skw(\nablastwo\R)=2\skw(\nablas\R\curvature\otimes\normal),
	\end{equation}
applied to $\nablas\R$ in \eqref{eq:nabla_R}.
\end{remark}
\begin{remark}
	\label{rmk:symmetry}
	While we wish the orientation of $\surface$ to be determined by a given normal field $\normal$, one could still exchange the directions $\e_1$ and $\e_2$, and consequently  $\kappa_1$ and $\kappa_2$, without affecting the curvature tensor in \eqref{eq:structure_laws_curvature}. The following transformations preserve  the orientation of $\surface$ and make the laws \eqref{eq:structure_laws_curvature} invariant,
	\begin{equation}
		\label{eq:symmetry_1}
		\e_1\to\e_2,\quad\e_2\to-\e_1,\quad\normal\to\normal,\quad\kappa_1\to\kappa_2,\quad\kappa_2\to\kappa_1,\quad\cv\to\cv.
	\end{equation}
	When accompanied by 
	\begin{equation}
		\label{eq:symmetry_2}
		\av_1\to\av_2,\quad\av_2\to-\av_1,
	\end{equation} 
	they also leave \eqref{eq:nabla_v} and \eqref{eq:integrability_conditions_rotation} unchanged.
\end{remark}

Combining \eqref{eq:integrability_conditions_rotation} and \eqref{eq:H_preserving_isometry}, we conclude that the existence of fields $(\av_1,\av_2,\av_3)$  that solve this (apparently over-determined) system of equations for a surface $\surface$ proves that the latter is a Bonnet surface and  provides the associated soft elastic deformations. 

For the moment, we set aside the issue of the compatibility of equations \eqref{eq:H_preserving_isometry} and \eqref{eq:integrability_conditions_rotation}, and we assume that there is a surface $\surface$ such that this combined system has a solution. We are interested in constructing the corresponding soft elastic deformations of $\surface$. 

To dispose of any overall uniform isometry that could be superimposed on the soft elastic deformations $\y$ we seek, we first select an \emph{anchor} point $\x_0$ on $\surface$ and require $\y(\x_0)=\x_0$; this disposes of possible translations. To dispose of rotations, we prescribe that the frames $\framev$ and $\framen$ coincide at $\x_0$. 

To construct $\surface^\ast$, we first integrate equations \eqref{eq:nabla_v}, which are compatible on any solution  of \eqref{eq:integrability_conditions_rotation}. To this end, we consider a path $\rv$ on $\surface$ joining $\x_0$ with $\x\in\surface$; we regard $\rv$ as a piece-wise smooth, one-parameter mapping $\sigma\mapsto\rv(\sigma)\in\surface$ and denote by $\rvd\in\tplane_{\rv(\sigma)}$ the derivative of $\rv$ with respect to the parameter $\sigma$, wherever it exists. Along the path $\rv(\sigma)$, the frame $\framev$ can be differentiated with respect to $\sigma$ wherever $\rv$ can, and equations \eqref{eq:nabla_v} acquire the compact form
\begin{equation}
	\label{eq:v_differential}
	\begin{pmatrix}
		\dot{\vv}_1\\
		\dot{\vv}_2\\
		\dot{\normal}^\ast
	\end{pmatrix}=
	\WW(\dot{\rv})
	\begin{pmatrix}
		\vv_1\\
		\vv_2\\
		\normal^\ast
	\end{pmatrix},
\end{equation}
where $\WW$ is the skew-symmetric matrix defined as
\begin{equation}
	\label{eq:W_matrix}
	\WW(\dot{\rv}):=
	\begin{pmatrix}
		0&\cv\cdot\rvd&-[\kappa_1(\e_1\cdot\rvd)+\av_2\cdot\rvd]\\
		-\cv\cdot\rvd&0&-[\kappa_2(\e_2\cdot\rvd)-\av_1\cdot\rvd]\\
		[\kappa_1(\e_1\cdot\rvd)+\av_2\cdot\rvd]&[\kappa_2(\e_2\cdot\rvd)-\av_1\cdot\rvd]&0
	\end{pmatrix}.
\end{equation}
Thus, by choosing appropriate paths $\rv$ on $\surface$, we reduce the problem of finding the frame $\framev$ that solves \eqref{eq:nabla_v} to the problem of solving a system of ordinary differential equations. Letting
\begin{equation}
	\label{eq:A_matrix}
	\AA(\rv):=\int\WW(\dot{\rv}(\sigma))\dd\sigma,
	\end{equation}
we can give the solution of \eqref{eq:v_differential} the form
\begin{equation}
	\label{eq:v_solution}
	\begin{pmatrix}
		\vv_1\\
		\vv_2\\
		\normal^\ast
	\end{pmatrix}(\x)=
	\ex^{\AA(\rv)}\left.
	\begin{pmatrix}
		\vv_1\\
		\vv_2\\
		\normal^\ast
	\end{pmatrix}\right|_{\x_0}=
	\ex^{\AA(\rv)}\left.
	\begin{pmatrix}
		\e_1\\
		\e_2\\
		\normal
	\end{pmatrix}\right|_{\x_0},
\end{equation}
on all paths for which
\begin{equation}
	\label{eq:commutation_condition}
	\AA(\rv)\WW(\dot{\rv})=\WW(\dot{\rv})\AA(\rv),
\end{equation}
see, for example \citep[p.\,38]{hurewicz:lectures}.

By Rodrigues's formula \citep[p.\,292]{marsden:introduction}, 
\begin{equation}
	\label{eq:A_representation}
	\ex^\AA=\mathbb{I}+\frac{1}{\alpha}\sin\alpha\AA+\frac{1}{\alpha^2}(1-\cos\alpha)\AA^2,
\end{equation}
where $\mathbb{I}$ is the identity matrix and $\alpha$ is related to the Frobenius norm $\|\AA\|$ of $\AA$ by
\begin{equation}
	\label{eq:alpha_definition}
	\alpha:=\frac{\|\AA\|}{\sqrt{2}}.
\end{equation}

\begin{remark}
	\label{rmk:luxury}
	Whenever we cannot cover $\surface$ with paths $\rv$ along which \eqref{eq:commutation_condition} applies, we shall not enjoy the luxury of representing explicitly the solution of \eqref{eq:v_differential} in the form \eqref{eq:v_solution}.
\end{remark}
Letting $\U=\proj$ in \eqref{eq:deformation_gradient}, as required for an isometry, and using \eqref{eq:R_curvature_frame}, we write
\begin{equation}
	\label{eq:nablas_y}
	\nablas\y=\vv_1\otimes\e_1+\vv_2\otimes\e_2,
\end{equation}
and by integrating again along the path $\rv$, we readily arrive at
\begin{equation}
	\label{eq:y_solution}
	\y(\x)=\x_0+\int[(\rvd\cdot\e_1)\vv_1+(\rvd\cdot\e_2)\vv_2]\dd\sigma,
\end{equation}
where $\vv_1$ and $\vv_2$ satisfy \eqref{eq:v_differential}. This is the formal solution for a soft elastic deformation of a Bonnet surface $\surface$; provided that \eqref{eq:y_solution} can be injectively extended on $\surface$, it delivers the material surface $\surface^\ast$ into which $\surface$ can be deformed with no energy expenditure.

\begin{remark}
\label{rmk:y_integrability}
It is an easy exercise to check that by \eqref{eq:structure_laws_curvature} and \eqref{eq:nabla_v} the tensor field in \eqref{eq:nablas_y} indeed obeys the integrability condition \eqref{eq:integrability_vector}.
	\end{remark}

\begin{remark}
	\label{rml:multiplicity}
	It should be clear 
	that each solution $(\av_1,\av_2,\av_3)$ of \eqref{eq:H_preserving_isometry} and \eqref{eq:integrability_conditions_rotation}, if it exists for a given $\surface$, generates a soft elastic deformation of $\surface$. We shall return later to the multiplicity of such solutions.
\end{remark} 
In the following section, we shall prove that the equations in the combined system \eqref{eq:H_preserving_isometry} and \eqref{eq:integrability_conditions_rotation} are indeed compatible for selected classes of surfaces, thus also confirming in the formalism adopted in this paper results known from  the classical literature on Bonnet surfaces. In Sect.~\ref{sec:applications}, we shall apply the method outlined in this section to construct families of soft elastic deformations of some Bonnet surfaces.
\begin{remark}
	\label{rmk:applications}
	It will become clearer in the following, but it is worth reminding the reader from the start that here we are \emph{not} interested in Bonnet surfaces that admit a \emph{discrete} number of soft elastic deformations, as they do not bring to life a ground state sufficiently rich in configurations to generate soft elasticity within  our theory for soft shells.
\end{remark}

\section{Bonnet Surfaces}\label{sec:bonnet}
The quest for Bonnet surfaces was initiated by Bonnet himself and has received important contributions, some even very recently, which we now briefly review.\footnote{A rather thorough account is given in the book by \citet[Chapt.~10]{jensen:surfaces}.} This is a classical topic, which in the literature on differential geometry of surfaces in three-dimensional space has a motivation quite different from ours. Before phrasing in our language the equations that govern the existence of Bonnet surfaces along with their continuum of soft elastic deformations, we shall make these differences clear in a number of technical remarks extracted from the literature.

The typical question that leads to Bonnet surfaces concerns uniqueness. Any isometric deformation of a surface $\surface$ is known to preserve the Gaussian curvature $K$ of $\surface$: is a non-trivial\footnote{That is, leaving rigid motions aside.} isometric deformation of $\surface$ uniquely determined by requiring that the mean curvature $H$ of $\surface$ is also preserved?
\begin{remark}
	\label{rmk:riemann}
	In Riemannian geometry, this question amounts to ask whether, given an abstract surface $\mathfrak{S}$ and having assigned a metric $\mathfrak{g}$ and a scalar function $H$ on $\mathfrak{S}$, there is a unique immersion of $\mathfrak{S}$ in three-dimensional space $\euclid$ with metric $\mathfrak{g}$ and mean curvature $H$. By a classical theorem of \citet{bonnet:memoire_II}, see also \citep[pp.\,249]{pressley:elementary} or \citep[p.\,70]{jensen:surfaces}, patches of surfaces in $\euclid$ with equal first and second fundamental forms may only differ by a rigid motion. Now, the first fundamental form is nothing but the metric, while the second fundamental form is a symmetric bilinear form with respect to the metric; the Gaussian and mean curvature are just the determinant and half the trace of the second fundamental form, the former being fully determined by the metric. Thus, a surface \emph{fails} to be a Bonnet surface if it is \emph{uniquely} determined by the metric and the trace of  the second fundamental form.
\end{remark}
We have already recalled in Remark~\ref{rmk:minimal_surfaces} that there exist minimal Bonnet surfaces; there are indeed many more. We follow \citep{bobenko:compact} in distinguishing the following categories of Bonnet surfaces.
\begin{enumerate}[label=(\arabic*)]
	\item \emph{Surfaces with constant mean curvature}. These surfaces, apart from plane and sphere, admit a one-parameter family of soft elastic deformations. The literature on global studies for constant mean curvature surfaces is vast; we refer the reader to the recent work of \citet{traizet:construction} as a mere token of liveness of this research field.
	\item \emph{Bonnet families}. These form a finite dimensional space of surfaces with non-constant mean curvature, each exhibiting a one-parameter family of soft elastic deformations. Local studies of these surfaces were conducted by \citet{cartan:couples} and \citet{chern:deformation};  global studies can be found in \citep{roussos:number}, \citep{roussos:global},  \citep{bobenko:bonnet}, and \citep[\S\,3]{bobenko:painleve}.
	\item \emph{Bonnet pairs}. These are isolated surfaces linked by a \emph{single} isometry that preserves their mean curvature. They may have either constant or non-constant mean curvature. Apart from the notable exception of \citep{bobenko:compact}, little is known about these surfaces. Sufficient conditions have been given for them to fail to exist by \citet{sabitov:isometric} and \citet{jensen:compact}. The members of a Bonnet pair, regarded as material soft shells, share the same elastic energy, but an energy barrier separates them, and no spontaneous transition between them can be expected.\footnote{A typical example, also recalled by \citet{sonnet:metric}, is the elastic \emph{eversion} of a shell, where two configurations sharing the same energy can be obtained one from the other through a costly \emph{snapping}.}
\end{enumerate}
\begin{remark}
	\label{rmk:helicoidal}
	\emph{Helicoidal} surfaces have long been known to exhibit all three types of Bonnet surfaces \citep{roussos:helicoidal}.\footnote{A helicoidal surface is produced by a uniform rigid rotation of a curve in three-dimensional space.}
\end{remark}
To abridge our lexicon, we introduce the following shorthand.
\begin{definition}
	\label{def:mode}
	Whenever there is a one-parameter family of soft elastic deformations of a \emph{Bonnet surface}, we shall say that this admits a \emph{soft mode}.\footnote{Thus somewhat interpreting as \emph{time} the parameter describing the family.}
\end{definition}
\begin{remark}
	\label{rmk:nirenberg}
	In the class of \emph{compact} surfaces (that are both closed and bounded), the quest for Bonnet surfaces must come to terms with the higher \emph{rigidity} expected in general in that class.\footnote{For example, for smooth compact surfaces that are also \emph{convex} is known that the only isometric motion, that is, a motion traversing only isometric surfaces, is a rigid motion. This is a result attributed to Blaschke and Weyl, as admirably recounted by \citet[pp.\,132-136]{nirenberg:lectures}.} \citet{lawson:mean} proved that  compact surfaces with non-constant mean curvature can at most exhibit Bonnet pairs and there are none with genus $g=0$. Until recently, it was not known whether there exist Bonnet pairs with genus $g\geqq1$. \citet{bobenko:compact} proved that compact Bonnet surfaces with $g=1$ do indeed exist and are uncountably many. They appear as \emph{twisted} tori, for which an enlightening illustration was provided by \cite{cutts:twisty} alongside a  popularized history of Bonnet surfaces.
	\end{remark}
\begin{remark}
	\label{rmk:genus}
	It is still not known whether there are compact Bonnet surfaces with genus $g>1$, neither whether there are compact Bonnet pairs with constant mean curvature.\footnote{To which the limitation in cardinality imposed by the theorem of \citet{lawson:mean} does not apply.} The latter might get across  Bonnet families with constant mean curvature: unlike the members in one and the same family, members in two distinct families might be paired by a soft elastic deformation, but not by a soft mode.
\end{remark}
\begin{remark}
	\label{rmk:revolution}
	Special classes of non-compact Bonnet families with non-constant mean curvature have also been explored. These include tangential developable surfaces \citep{roussos:tangential}, surfaces of \emph{revolution}, whose profile curve satisfies a fourth-order ordinary differential equation, and \emph{flat} surfaces, with vanishing Gaussian curvature $K\equiv0$ \citep{roussos:principal}. It was proved by \citet{colares:isometric} that surfaces with constant Gaussian curvature $K\neq0$ \emph{cannot} be Bonnet surfaces. 
\end{remark}
We now present our criterion for the existence of a soft mode.

\subsection{Soft Mode Criterion}
The equations that need to be solved to identify a soft mode are \eqref{eq:H_preserving_isometry} and \eqref{eq:integrability_conditions_rotation}. By \eqref{eq:H_preserving_isometry_2} and \eqref{eq:H_preserving_isometry_3}, we can represent the vectors $\av_1$ and $\av_2$ that represent $\WW$ in the frame $\framen$ as
\begin{equation}
	\label{eq:alphas}
	\av_1=\alpha_1\e_1+\alpha_2 \e_2,\qquad\av_2=\alpha_2\e_1-\alpha_1\e_2.
\end{equation}
With these, equation \eqref{eq:integrability_conditions_rotation_1} becomes
\begin{equation}
	\label{eq:alphas_equation}
	\alpha_2(\kappa_2-\kappa_1)=\alpha_1^2+\alpha_2^2,
\end{equation}
so that we can write
\begin{equation}
	\label{eq:alphas_representation}
	\alpha_1=\tfrac12(\kappa_2-\kappa_1)\sin2\vt,\qquad\alpha_2=\tfrac12(\kappa_2-\kappa_1)(1-\cos2\vt),
\end{equation}
where $\vt\in[0,\pi]$ is a field to be determined. Clearly, for $\vt\in\{0,\pi\}$, all $\av$ vectors vanish and the polar rotation $\R$ becomes uniform.
\begin{remark}
	\label{rmk:phi}
	The angle $\vt$ has a transparent geometric interpretation. To see this, we first make use of \eqref{eq:alphas} and \eqref{eq:alphas_representation} in \eqref{eq:nabla_v} to write $\nablas\normal^\ast$ in the form
\begin{equation}
\label{eq:nabla_normal_ast_rewritten}
\begin{split}
\nablas\normal^\ast=&\left[H-\tfrac12(\kappa_2-\kappa_1)\cos2\vt\right]\vv_1\otimes\e_1+\left[H+\tfrac12(\kappa_2-\kappa_1)\cos2\vt\right]\vv_2\otimes\e_2\\
-&\tfrac12(\kappa_2-\kappa_1)\sin2\vt(\vv_2\otimes\e_1+\vv_1\otimes\e_2).
\end{split}	
\end{equation}
On the other hand, by the chain rule, 
\begin{equation}
	\label{eq:nabla_normal_ast_alternate}
	\nablas\normal^\ast=(\nablast\normal^\ast)(\nablas\y)=(\kappa_1\e_1^\ast\otimes\e_1^\ast+\kappa_2\e_2^\ast\otimes\e_2^\ast)(\vv_1\otimes\e_1+\vv_2\otimes\e_2),
\end{equation}
where we have used \eqref{eq:deformation_gradient} and \eqref{eq:U} to express $\nablas\y$ and the fact that the curvature tensor $\nablast\normal^\ast$ of $\surface^\ast$ has the same eigenvalues as the curvature tensor $\nablas\normal$ of $\surface$ and eigenvectors $(\e_1^\ast,\e_2^\ast)$. By representing these latter as
\begin{equation}
	\label{eq:e_star}
	\e_1^\ast=\cos\vp\vv_1+\sin\vp\vv_2,\qquad \e_2^\ast=-\sin\vp\vv_1+\cos\vp\vv_2,
\end{equation}
and inserting \eqref{eq:e_star} into \eqref{eq:nabla_normal_ast_alternate}, simple calculations show that equations \eqref{eq:nabla_normal_ast_alternate} and \eqref{eq:nabla_normal_ast_rewritten} agree if and only if
\begin{equation}
\label{eq:theta=phi}
	\vp=\vt.
\end{equation}
Thus, $\vt$ is the rotation angle of the principal frame $(\e_1^\ast,\e_2^\ast)$ of $\surface^\ast$ relative to the materially convected principal frame $(\vv_1,\vv_2)$ of $\surface$.
\end{remark}
We shall now assume that $\surface$ has \emph{no} umbilical points, that is,
\begin{equation}
	\label{eq:umbilics}
	\kappa_1\neq\kappa_2.
\end{equation}
\begin{remark}
	\label{rmk:umbilics}
	The classical local theory of Bonnet surfaces has been developed under this assumption. As remarked by \citet{roussos:global}, this restriction does not engender a real global difference, except in essentially \emph{one} singular case, where the set of umbilical points of $\surface$ consists of a single point, which is also the only critical point of the mean curvature $H$. We shall leave this exceptional case aside.
\end{remark}
By use of \eqref{eq:kappa_c_equations}, \eqref{eq:alphas}, and \eqref{eq:alphas_representation}, equations \eqref{eq:integrability_conditions_rotation_bis_2} and \eqref{eq:integrability_conditions_rotation_bis_3} can be given the following form,
\begin{equation}
	\label{eq:theta_system}
	\begin{dcases}
		\nablas\vt\cdot\e_1=\frac{1}{\kappa_2-\kappa_1}\{\sin2\vt\nablas H\cdot\e_1+(1-\cos2\vt)\nablas H\cdot\e_2\},\\
		\nablas\vt\cdot\e_2=\frac{1}{\kappa_2-\kappa_1}\{(1-\cos2\vt)\nablas H\cdot\e_1-\sin2\vt\nablas H\cdot\e_2\}.
	\end{dcases}
\end{equation}

	\begin{remark}
		\label{rmk:symmetry_bis}
		By \eqref{eq:alphas}, the symmetry transformations \eqref{eq:symmetry_1} and \eqref{eq:symmetry_2} affect $\alpha_1$ and $\alpha_2$ as follows,
		\begin{equation}
			\label{eq:symmetry_3}
			\alpha_1\to-\alpha_1,\quad\alpha_2\to-\alpha_2,
		\end{equation}
		and so, by \eqref{eq:alphas_representation}, $\vt$ is invariant, and so is the system \eqref{eq:theta_system}.
\end{remark}
Letting 
\begin{equation}
	\label{eq:tau_definition}
\tau:=\tan\vt,
\end{equation}
the system \eqref{eq:theta_system} can also be written more compactly as
\begin{equation}
	\label{eq:tau_system}
	\begin{dcases}
		\tau_{,1}=\frac{\sgn(\kappa_2-\kappa_1)}{\sqrt{H^2-K}}(\tau H_{,1}+\tau^2H_{,2}),\\
		\tau_{,2}=\frac{\sgn(\kappa_2-\kappa_1)}{\sqrt{H^2-K}}(\tau^2H_{,1}-\tau H_{,2}),
	\end{dcases}
\end{equation}
where we have set
\begin{equation}
	\label{eq:tau_notation}
	\tau_{,i}:=\nablas\tau\cdot\e_i\quad\text{and}\quad H_{,i}:=\nablas H\cdot\e_i.
\end{equation}

Surfaces for which the first-order partial differential equations in \eqref{eq:theta_system} admit solutions are the Bonnet surfaces we are interested in: they have a soft mode, parameterized in the single constant of integration that  these solutions possess. For every solution $\vt$ of \eqref{eq:theta_system},  we determine $\av_1$ and $\av_2$ via \eqref{eq:alphas_representation} and \eqref{eq:alphas}; inserting these vector fields into \eqref{eq:W_matrix}, and eventually into \eqref{eq:v_solution} and \eqref{eq:y_solution}, with the aid of \eqref{eq:A_matrix} and \eqref{eq:A_representation}, we determine the corresponding mode.  
\begin{remark}
	\label{rmk:identity}
	It is perhaps worth recalling that $\vt\equiv0$ is a solution of \eqref{eq:theta_system} for all surfaces $\surface$. This special solution entails that $\av_1=\av_2=\av_3=\zero$, and so the corresponding polar rotation $\R$ is uniform on the whole of $\surface$. By our choice of \emph{anchoring} $\surface$ and $\surface^\ast$ at a point $\x_0$, where both $\y(\x_0)=\x_0$ and $\vv_i=\e_i$, the deformation $\y$ in \eqref{eq:y_solution} would simply be the identity and its polar rotation $\R$ would be $\Id$.
We shall only seek solutions of \eqref{eq:theta_system} that do not vanish identically.
\end{remark} 

\begin{remark}
	\label{rmk:path_independence}
	The representations for $\av_1$ and $\av_2$ in \eqref{eq:alphas} and \eqref{eq:alphas_representation} along with the integrability conditions \eqref{eq:integrability_conditions_rotation} allow us to identify a general sufficient condition for the matrix $\AA(\rv)$ to be independent of the integration path $\rv$, thus being only determined by the  points joined by $\rv$. This condition is
	\begin{equation}
		\label{eq:c=0}
		\cv\equiv\zero,
	\end{equation}
	where $\cv$ is the spin connector of the principal curvature frame $\framen$. As recalled in Remark~\ref{rmk:c=0}, \eqref{eq:c=0} requires that the Gaussian curvature $K$ of $\surface$ vanishes. The proof of  this condition is given in Appendix~\ref{sec:c=0}, where it is also shown that it falls short of being a necessary condition for the path-independence of $\AA$, whereas $K\equiv0$ always is.
\end{remark}

\subsection{Special Solutions}
Surfaces with constant mean curvature are immediately seen to provide solutions to \eqref{eq:theta_system}: these are $\vartheta\equiv\vartheta_0$, with $\vt_0$ constant. Varying $\vt_0$, we can construct a soft mode by applying the method illustrated in Sect.~\ref{sec:soft}. An explicit example of this construction will be given in Sect.~\ref{sec:applications} for a circular cylinder.
\begin{remark}
	\label{rmk:theta_non_constant}
	It is easy to show that if \eqref{eq:umbilics} applies then the system \eqref{eq:tau_system} has no constant solution $\vt$, unless $H$ is constant. Thus, if $\surface$ is a Bonnet surface with non-constant mean curvature with a soft mode, then by Remark~\ref{rmk:phi} the principal frame of $\surface^\ast$ makes a non-constant angle with the materially convected principal frame of $\surface$. 
\end{remark}

Next we study flat surfaces, both built on spirals.

\subsubsection{Spiral Cylinders}
Consider a cylinder $\spicyl$ generated by translating along the unit vector $\e_z$ of a Cartesian frame $\framec$ a plane curve $\rv_0(s)$ lying on the $(x,y)$ plane,
\begin{equation}
	\label{eq:r_cylinder}
	\rv(s,z)=\rv_0(s)+z\e_z,
\end{equation}
where $s$ is the arc-length parameter of the curve $\rv_0$. The principal directions of curvature are along $\e_1=\e_z$ with $\kappa_1=0$ and along $\e_2=\rv_0'$ with $\kappa_2=\kappa(s)$, where a prime denotes differentiation with respect to $s$ and $\kappa(s)$ is the curvature of $\rv_0$.

Equations \eqref{eq:theta_system} readily become the system
\begin{equation}\label{eq:theta_system_cylider}
\begin{dcases}
	\partial_z\vt=\frac12\frac{\kappa'}{\kappa}(1-\cos2\vt),\\
	\partial_s\vt=-\frac12\frac{\kappa'}{\kappa}\sin2\vt,
\end{dcases}
\end{equation}	
for which the integrability condition $\partial_{zs}^2\vt=\partial_{sz}^2\vt$ delivers the equation
\begin{equation}
	\label{eq:integrability_cylinder}
	\frac12\left[\left(\frac{\kappa'}{\kappa}\right)'-\left(\frac{\kappa'}{\kappa}\right)^2\right](1-\cos2\vt)=0.
\end{equation}
Letting $\vt\notin\{0,\pi\}$, to rule out trivial solutions, we see that $\spicyl$ is a Bonnet surface whenever
\begin{equation}
	\label{eq:integrability_cylinder_non_trivial}
	\left(\frac{\kappa'}{\kappa}\right)'-\left(\frac{\kappa'}{\kappa}\right)^2=0,
\end{equation}
the solution of which is 
\begin{equation}
	\label{eq:integrability_cylinder_solution}
	\kappa(s)=\frac{1}{\mu s+\lambda},
\end{equation}
with both $\mu$ and $\lambda$ arbitrary constants. Use of this in \eqref{eq:tau_system} leads us to the system
\begin{equation}
	\label{eq:tau_system_cylinder}
	\begin{dcases}
		\partial_z\tau=-\frac{\mu\tau^2}{\mu s+\lambda},\\
		\partial_s\tau=\frac{\mu\tau}{\mu s+\lambda},
	\end{dcases}
\end{equation}
which is solved by
\begin{equation}
	\label{eq:tau_system_cylinder_solution}
	\tau=\frac{\mu s+\lambda}{\mu z+c_0},
\end{equation}
where $c_0$ is a constant of integration.

Equation \eqref{eq:integrability_cylinder_solution} describes a family of logarithmic spirals in the plane. For each of the corresponding  spiral cylinders $\spicyl$, \eqref{eq:tau_system_cylinder_solution} describes its soft mode, parameterized in $c_0$, that recovers $\spicyl$ for $c_0\to\pm\infty$. This mode will be shown shortly below in Sect.~\ref{sec:applications}.

\subsubsection{Spiral Cones}
Let now $\cone$ be a generalized cone projected from the center of the unit sphere $\sphere$; it can be represented as
\begin{equation}
	\label{eq:spiral_cone_representation}
	\rv(s,t)=t\e(s),
\end{equation}
where $t\geqq0$ is the arc-length parameter running along the straight lines emanating from the origin (sitting at $t=0$) and $s$ is the arc-length parameter of the curve $\e(s)$ where $\cone$ and $\sphere$ intersect. Setting $\e_1=\e(s)$  and $\e_2=\e'(s)$, where $\e'$ is the unit tangent along the curve $\e(s)$, the curvature tensor of $\cone$ reads as
\begin{equation}
	\label{eq:cone_curvature_tensor}
	\nablas\normal=\frac{1}{t}\kg(s)\e_2\otimes\e_2,
\end{equation}
where $\kg(s):=\e''\cdot\e'\times\e$ is the geodesic curvature of $\e(s)$ on $\sphere$, so that $\kappa_1=0$ and $\kappa_2=\kg(s)/t$.

Letting $\vt=\vt(s,t)$, we easily see that 
\begin{equation}
	\label{eq:cone_theta_differentials}
	\nablas\vt\cdot\e_1=\partial_t\vt,\qquad\nablas\vt\cdot\e_2=\frac1t\partial_s\vt
\end{equation}
and give equations \eqref{eq:theta_system} the following form
\begin{equation}
	\label{eq:theta_system_cone}
	\begin{dcases}
		\partial_t\vt=\frac12\frac1t\left\{\frac{\kg'}{\kg}(1-\cos2\vt)-\sin2\vt\right\},\\
		\partial_s\vt=-\frac12\left\{\frac{\kg'}{\kg}\sin2\vt+(1-\cos2\vt)\right\},
	\end{dcases}
\end{equation} 
which admit non-trivial solutions if and only if
\begin{equation}
	\label{eq:integrability_cone_non_trivial}
	\left(\frac{\kg'}{\kg}\right)'-\left(\frac{\kg'}{\kg}\right)^2-1=0.
\end{equation}
Equation \eqref{eq:integrability_cone_non_trivial} is solved by
\begin{equation}
\label{eq:integrability_cone_solution}
	\kg(s)=\mu\sec(s+\lambda),
\end{equation}
where both $\mu$ and $\lambda$ are arbitrary constants. These identify a family of spherical spirals that project a conical Bonnet surface $\cone$.

By inserting \eqref{eq:integrability_cone_solution} into \eqref{eq:tau_system} and solving for $\tau$, we arrive at
\begin{equation}
	\label{eq:tau_system_cone_solution}
	\tau=\frac{\cos(s+\lambda)}{\sin(s+\lambda)+c_0t}.
\end{equation}
Thus, for any given conical  Bonnet surface $\cone$, its soft mode is described by the parameter $c_0$ and $\cone$ is recovered in the limit as $c_0\to\pm\infty$.
\begin{remark}
	\label{rmk:rescaling}
	Either a rescaling or a shift in the origin of the arc-length $s$ in both \eqref{eq:integrability_cylinder_solution} and \eqref{eq:integrability_cone_solution} would affect neither $\spicyl$ nor $\cone$. Thus, there is a \emph{single} parameter describing all spiral cylinders or cones that are Bonnet surfaces.
\end{remark}
\begin{remark}
	\label{rmk:roussos}
	Both spiral cylinders and cones were identified as Bonnet surfaces by \citet{roussos:principal}, who first found and solved equations \eqref{eq:integrability_cylinder_non_trivial} and \eqref{eq:integrability_cone_non_trivial}. He also showed that  non-trivial isometries preserving the mean curvature can be established between spiral cylinders and spiral cones. That is, spiral cylinders and cones can also be Bonnet pairs. These results of \citet{roussos:principal} are neatly recounted and beautifully illustrated by \citep[pp.\,304--312]{jensen:surfaces}, see in particular their Prop.~10.16 for an effective summary.
\end{remark}
In the following section, we shall construct explicitly soft modes of both circular and spiral cylinders. While the former will be seen to generate \emph{twisted} circular cylinders, the latter will generate spiral cones.

\section{Soft Modes}\label{sec:applications}
Here we illustrate the method presented in Sect~\ref{sec:soft} to construct the soft modes of selected classes of Bonnet surfaces that meet the criterion given in Sect.~\ref{sec:bonnet}.

\subsection{Circular Cylinders}\label{sec:cylinder}
A circular straight cylinder $\cyl$ of radius $R$ and height $h$ satisfies equations \eqref{eq:theta_system} with $\vt\equiv\vt_0$ because it is a surface with constant mean curvature $H=1/2R$.

Let the axis of $\cyl$ be along the unit vector $\e_z$ of a Cartesian frame $\framec$. In cylindrical coordinates $(\phi,z)$, $\cyl$ is described by
\begin{equation}
	\label{eq:cylinder_representation}
	\rv(\phi,z)=R\e_r(\phi)+z\e_z,\qquad 0\leqq\phi\leqq2\pi,\quad 0\leqq z\leqq h.
\end{equation}
and the principal frame $\framen$ is constituted by
\begin{equation}
	\label{eq:cylinder_frame}
	\begin{cases}
		\e_1=\e_\phi=-\sin\phi\e_x+\cos\phi\e_y,\\
		\e_2=\e_z,\\
		\normal=\e_r=\cos\phi\e_x+\sin\phi\e_y,
	\end{cases}
\end{equation} 
with principal curvatures
\begin{equation}
	\label{eq:cylinder_curvatures}
	\kappa_1=\frac1R\quad\text{and}\quad\kappa_2=0.
\end{equation}

It is immediately seen (since $\nablas\e_z=\zero$) that for the frame $\framen$ the spin connector $\cv=\zero$. Moreover, it follows from \eqref{eq:alphas_representation} and \eqref{eq:alphas} that the vectors $\av_1$ and $\av_2$ are given by
\begin{equation}
	\label{eq:cylinder_a_vectors}
	\begin{cases}
		\av_1=-\frac{1}{2R}\{\sin2\vt_0\e_\phi+(1-\cos2\vt_0)\e_z\},\\
		\av_2=-\frac{1}{2R}\{(1-\cos2\vt_0)\e_\phi-\sin2\vt_0\e_z\}.
	\end{cases}
\end{equation}
Thus, the non-vanishing entries of the matrix in \eqref{eq:W_matrix} are
\begin{equation}
	\label{eq:cylinder_W}
	\WW_{13}=-\tfrac1R\e_\phi\cdot\dot{\rv}-\av_2\cdot\dot{\rv},\qquad\WW_{23}=\av_1\cdot\dot{\rv},
\end{equation}
where, by \eqref{eq:cylinder_representation},
\begin{equation}
	\label{eq:cylinder_r_dot}
	\dot{\rv}=R\dot{\phi}\e_\phi+\dot{z}\e_z.
\end{equation}
By Remark~\ref{rmk:path_independence}, $\AA$ is path-independent, as $\cv=\zero$;
an elementary integration delivers its non-vanishing entries in the following form,
\begin{equation}
	\label{eq:cylinder_A}
	\begin{cases}
		\AA_{13}(\phi,z)=-\frac12\phi(1+\cos2\vt_0)-\frac{z}{2R}\sin2\vt_0,\\
		\AA_{23}(\phi,z)=-\frac12\phi\sin2\vt_0-\frac{z}{2R}(1-\cos2\vt_0),
	\end{cases}
\end{equation}	
with the prescription that $\AA(0,0)=0$. Thus,
$\alpha$ in \eqref{eq:alpha_definition} acquires the explicit form
\begin{equation}
	\label{eq:cylinder_alpha}
\alpha=\sqrt{\tfrac12\left(\phi^2-\frac{z^2}{R^2}\right)\cos2\vt_0+ \frac{\phi z}{R}\sin2\vt_0+\tfrac12\left(\phi^2+\frac{z^2}{R^2}\right)}.
\end{equation}
Moreover, the commutation condition \eqref{eq:commutation_condition} here reduces to a single equation,
\begin{equation}
	\label{eq:commutation_condition_cylinder}
	\AA_{13}\WW_{23}=\AA_{23}\WW_{13},
\end{equation}
which, by \eqref{eq:cylinder_W} and \eqref{eq:cylinder_A}, turns out to be identically satisfied on all paths $\rv$. Thus the exponential representation \eqref{eq:v_solution} can be freely employed.

For any given $\vt_0\in(0,\pi)$, making use of \eqref{eq:cylinder_A} and \eqref{eq:cylinder_alpha} in \eqref{eq:v_solution} we obtain the moving frames $\framev$ as functions of $(\phi,z)$ on $\cyl$, and from these, performing the path integral in \eqref{eq:y_solution}, we determine the mapping $\y$ that describes the surface $\cyl^\ast(\vt_0)$ reached by the corresponding soft mode of $\cyl$.

\begin{figure}[] 
	\centering
	\begin{subfigure}[c]{2.2cm}
		\centering
		\includegraphics[width=2.2cm]{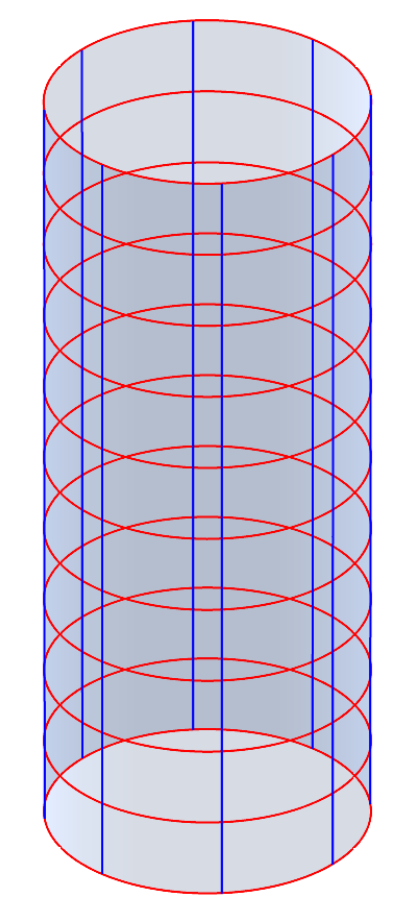}
		\caption{$\vt_0=0$}
		\label{fig:theta_0}
	\end{subfigure}
	\qquad
	\begin{subfigure}[c]{3cm}
		\centering
		\includegraphics[width=3cm]{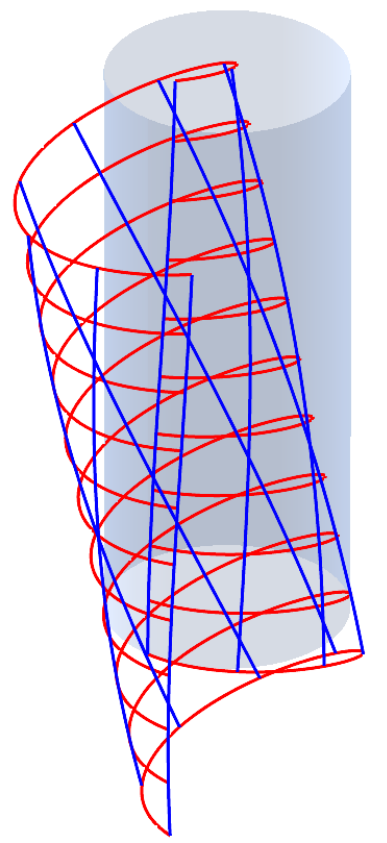}
		\caption{$\vt_0=\frac{\pi}{12}$}
		\label{fig:theta_1Pio12}
	\end{subfigure}
	\qquad
	\begin{subfigure}[c]{3cm}
		\centering
		\includegraphics[width=3cm]{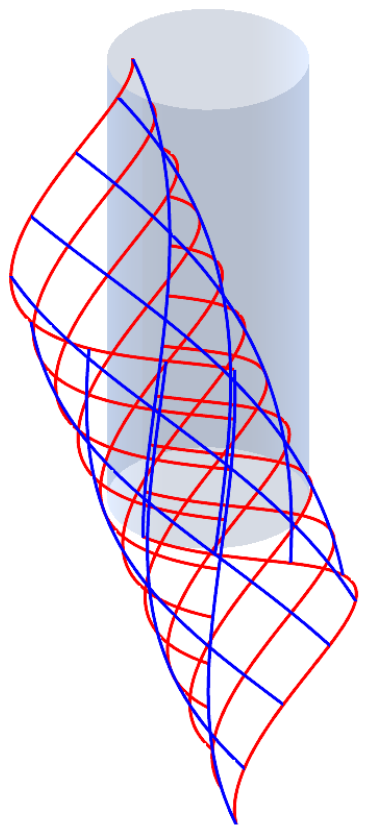}
		\caption{$\vt_0=\frac{2\pi}{12}$}
		\label{fig:theta_2Pio12}
	\end{subfigure}
	\qquad
	\begin{subfigure}[c]{3.4cm}
		\centering
		\includegraphics[width=3.4cm]{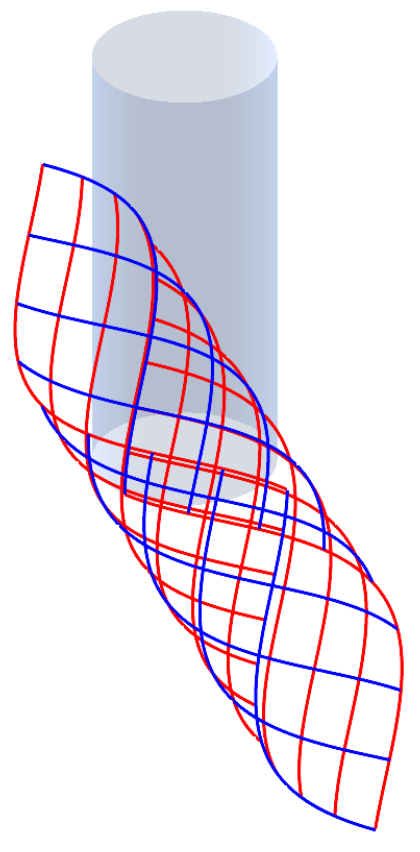}
		\caption{$\vt_0=\frac{3\pi}{12}$}
		\label{fig:theta_3Pio12}
	\end{subfigure}
	\\
	\begin{subfigure}[c]{4.3cm}
		\centering
		\includegraphics[width=4.3cm]{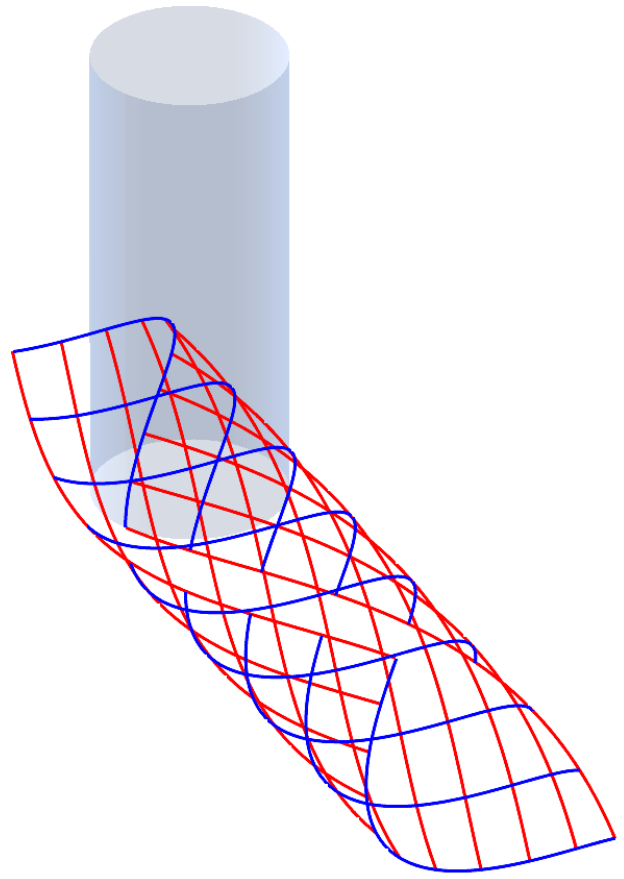}
		\caption{$\vt_0=\frac{4\pi}{12}$}
		\label{fig:theta_4Pio12}
	\end{subfigure}
	\qquad
	\begin{subfigure}[c]{4.3cm}
		\centering
		\includegraphics[width=4.3cm]{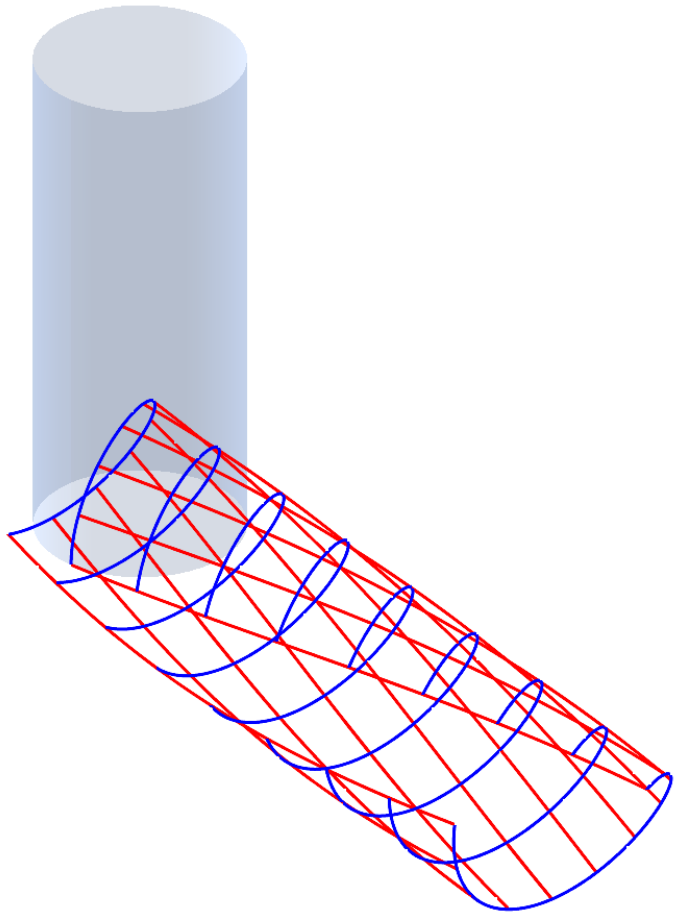}
		\caption{$\vt_0=\frac{5\pi}{12}$}
		\label{fig:theta_5Pio12}
	\end{subfigure}
	\qquad
	\begin{subfigure}[c]{4.6cm}
		\centering
		\includegraphics[width=4.6cm]{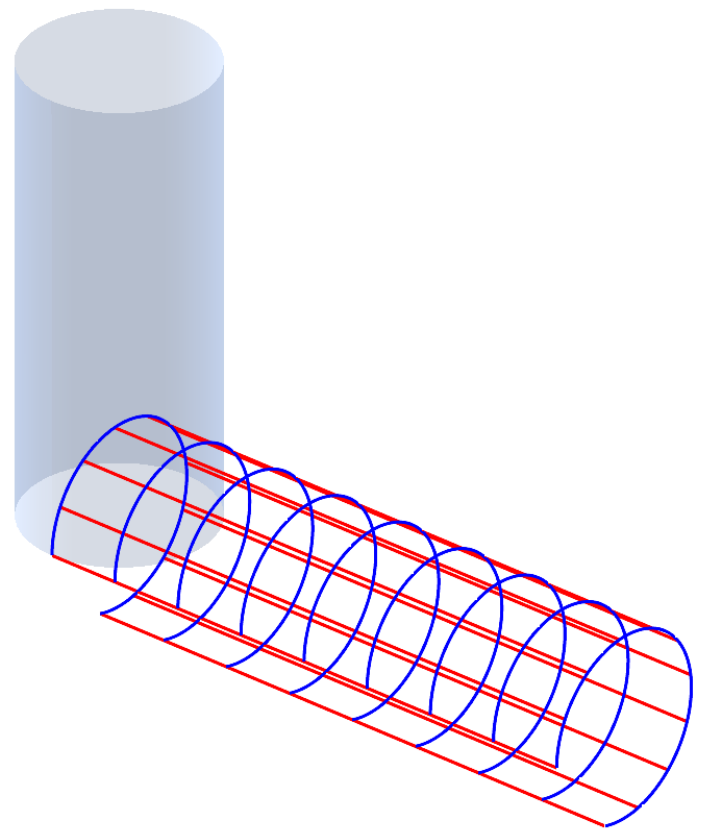}
		\caption{$\vt_0=\frac{6\pi}{12}$}
		\label{fig:theta_6Pio12}
	\end{subfigure}
	\\
	\begin{subfigure}[c]{4.6cm}
		\centering
		\includegraphics[width=4.6cm]{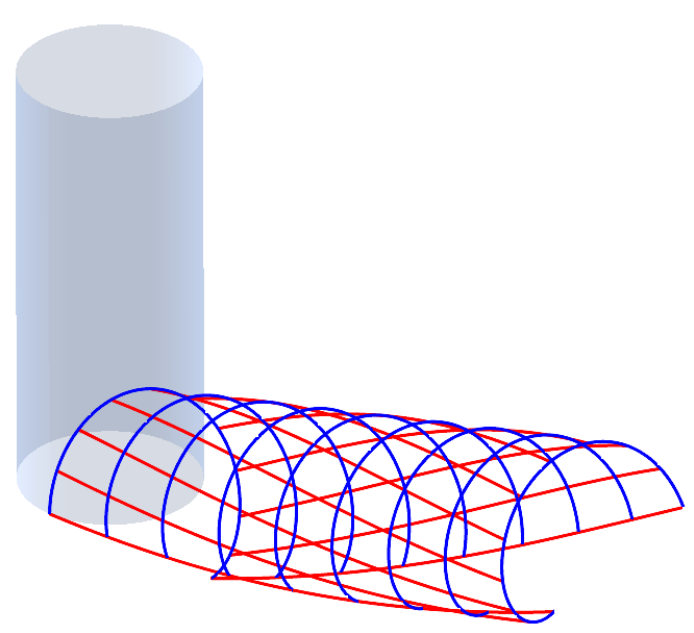}
		\caption{$\vt_0=\frac{7\pi}{12}$}
		\label{fig:theta_7Pio12}
	\end{subfigure}
	\qquad
	\begin{subfigure}[c]{4.2cm}
		\centering
		\includegraphics[width=4.2cm]{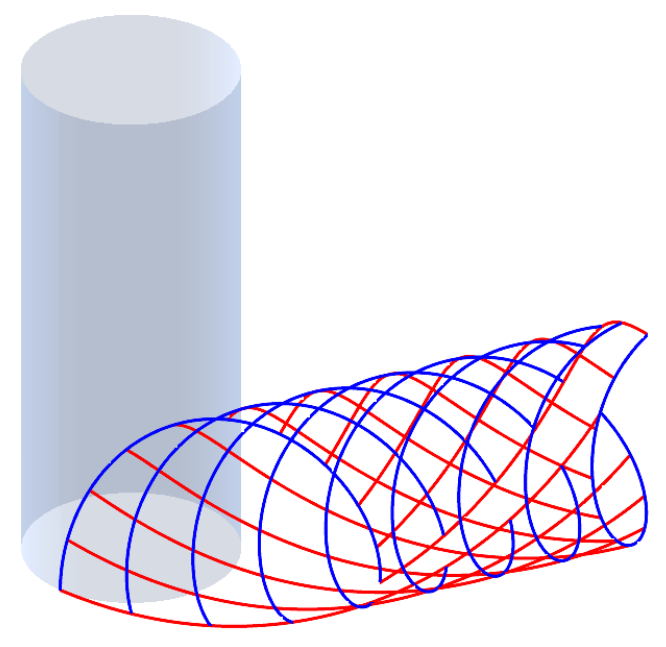}
		\caption{$\vt_0=\frac{8\pi}{12}$}
		\label{fig:theta_8Pio12}
	\end{subfigure}
	\qquad
	\begin{subfigure}[c]{4.2cm}
		\centering
		\includegraphics[width=4.2cm]{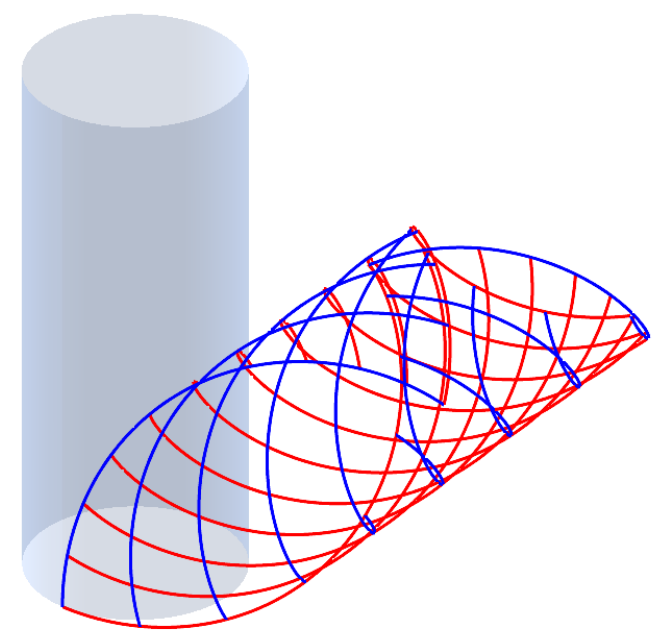}
		\caption{$\vt_0=\frac{9\pi}{12}$}
		\label{fig:theta_9Pio12}
	\end{subfigure}
	\\
	\begin{subfigure}[c]{3.5cm}
		\centering
		\includegraphics[width=3.5cm]{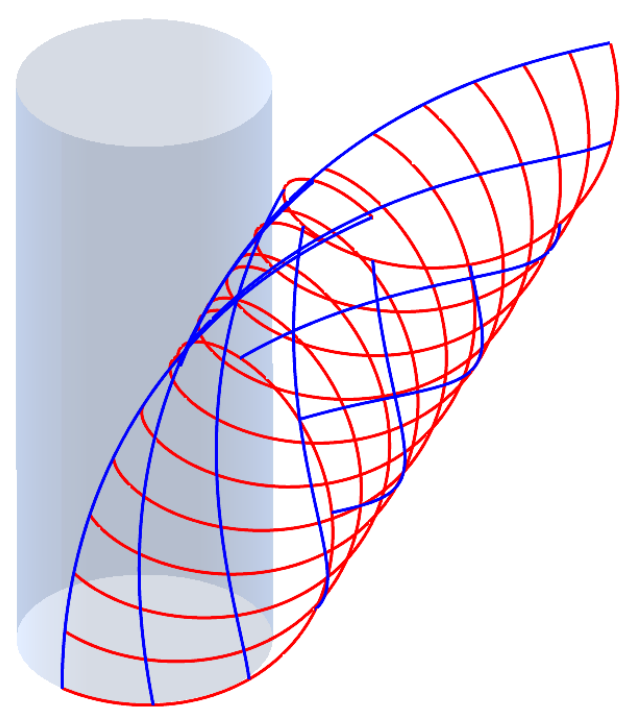}
		\caption{$\vt_0=\frac{10\pi}{12}$}
		\label{fig:theta_10Pio12}
	\end{subfigure}
	\qquad
	\begin{subfigure}[c]{2.5cm}
		\centering
		\includegraphics[width=2.5cm]{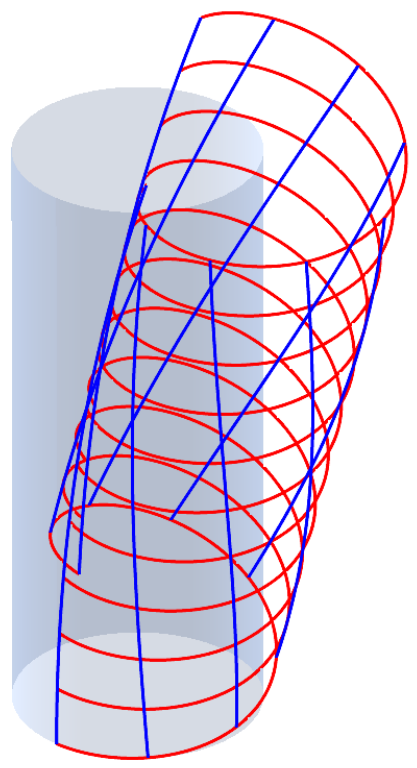}
		\caption{$\vt_0=\frac{11\pi}{12}$}
		\label{fig:theta_11Pio12}
	\end{subfigure}
	\caption{Configurations $\cyl^\ast(\vt_0)$ traversed by the soft mode of a straight circular cylinder $\cyl$, depicted in the background and described by \eqref{eq:cylinder_representation}. Here the radius of $\cyl$ is $R=1$ and its height is $h=5$.}
	\label{fig:cylinder_circular}
\end{figure}

We scale all lengths to $R$, so that $R=1$, and we choose the anchor point $\x_0=\e_x$. For example, with these choices and for $\vt_0=\frac{\pi}{2}$, we calculate
\begin{equation}
	\label{eq:cylinder_pi02}
	\y(\phi,z)=\cos z\e_x+\phi\e_y+\sin z\e_z,
\end{equation}
which represents a cylinder of radius $R=1$ and axis along $\e_y$. This deformation maps all meridians of $\cyl$ into parallels of $\cyl^\ast(\frac{\pi}{2})$ and all parallels of the former into meridians of the latter (see Fig.~\ref{fig:theta_6Pio12}).

Similarly, for $\vt_0=\frac{\pi}{4}$, we obtain
\begin{equation}
	\label{eq:cylinder_pio4}
	\begin{split}
	\y(\phi,z)=&\cos\left(\frac{\phi+z}{\sqrt{2}}\right)\e_x+\left\{\frac12(\phi-z)+\frac{1}{\sqrt{2}}\sin\left(\frac{\phi+z}{\sqrt{2}}\right)\right\}\e_y\\
	+&\left\{\frac12(z-\phi)+\frac{1}{\sqrt{2}}\sin\left(\frac{\phi+z}{\sqrt{2}}\right)\right\}\e_z.
	\end{split}
\end{equation}

Several configurations $\cyl^\ast(\vt_0)$ traversed by the soft mode of $\cyl$ are depicted in Fig.~\ref{fig:cylinder_circular}.
It is known that they all cover cylinders of radius $R=1$ \citep{colares:isometric} without preserving the axis of $\cyl$.\footnote{An isometric deformations of any small patch of a right circular cylinder is a rigid motion only when it preserves the axis of the cylinder \citep{wolf:surfaces}.}

Figure~\ref{fig:theta_0} shows the reference cylinder  $\cyl$ of height $h=5$, decorated with its parallels and meridians (red and blue, respectively). All other panels in Fig~\ref{fig:cylinder_circular} show $\cyl^\ast(\vt_0)$ through the net of deformed parallels and meridians, with the reference $\cyl$ in the background. As expected, see Remark~\ref{rmk:phi}, the deformed parallels and meridians, which are tangent to $\vv_1$ and $\vv_2$, are not lines of principal curvature (except for $\vt_0=\frac{\pi}{2}$).

To realize the soft mode illustrated in Fig.~\ref{fig:cylinder_circular}, one should imagine to cut $\cyl$ along the meridian through the anchor point and let the adjoining sides separate. This, for example, is what allows the meridians at $\phi=0$ and $\phi=2\pi$ to become the delimiting parallels of $\cyl^\ast(\frac{\pi}{2})$ sitting at the distance $2\pi$ from one another in Fig.~\ref{fig:theta_6Pio12}. All parallels of $\cyl^\ast(\frac{\pi}{2})$ fail to be closed because we have chosen $h<2\pi$; for $h>2\pi$, they would be wrapped upon themselves, giving rise to interpenetration. Other instances of interpenetration occur in Fig.~\ref{fig:cylinder_circular}; they should strictly be banned from mechanics, but we tolerate them for sake of illustration.

The general feature revealed by Fig.~\ref{fig:cylinder_circular} is that the soft mode of a straight circular cylinder is realized by \emph{twisting} its parallels and meridians.

\subsection{Spiral Cylinders}\label{sec:spiral}
We consider a specific spiral cylinder $\spicyl$ as in \eqref{eq:r_cylinder}, generated by the logarithmic spiral defined in parametric form by
\begin{equation}
r_0(\phi)=\frac{1}{\sqrt{2}} e^\phi,\quad -\infty<\phi<\infty,
\end{equation}
and we measure its arclength $s$ starting from the origin as
\begin{equation}
s(\phi)=\int_{-\infty}^\phi\sqrt{\left[r_0(\varphi)\right]^2+\left[r_0'(\varphi)\right]^2}d\varphi=e^\phi
\end{equation}
so that $\phi=\ln s$ and $r_0(s)=\frac{s}{\sqrt{2}}$. Thus, in Cartesian coordinates, we have
\begin{equation}
\rv_0(s)=r_0(s)\e_r=\frac{s}{\sqrt{2}}\left(
\cos(\ln(s))\e_x+\sin(\ln(s))\e_y
\right).
\end{equation}

The unit tangent $\tangent$ to the spiral is
\begin{equation}
\tangent(s)=\rv_0'(s)=\frac{1}{\sqrt{2}}\left(\e_r+\e_\phi\right)
\end{equation}
with
\begin{equation}
\e_\phi=\left(-\sin(\ln(s))\e_x+\cos(\ln(s))\e_y\right).
\end{equation}
and so with $\e_1=\e_z$ and $\e_2=\tangent$ we have
\begin{equation}\label{eq:spiralNormal}
 \normal=\e_z\times\tangent=\frac{1}{\sqrt{2}}\left(\e_\phi-\e_r\right).
\end{equation}

The modulus of the spiral's curvature is
\begin{equation}
\vert\kappa(s)\vert=\vert \tangent'(s)\vert=\frac{1}{s}.
\end{equation}
In order to determine the correct sign of the curvature, note that the second equation of \eqref{eq:structure_laws_curvature}, when applied
to $\tangent$, here becomes
\begin{equation}\label{eq:MinusFrenet}
\tangent'=-\kappa\normal.
\end{equation}
Thus, with the normal \eqref{eq:spiralNormal}, for \eqref{eq:MinusFrenet} to
be satisfied we need
\begin{equation}\label{eq:spiralCurvature}
\kappa(s)=-\frac{1}{s},
\end{equation}
and moreover we easily find that  $\cv=\zero$.

The form of the curvature \eqref{eq:spiralCurvature} shows that our choices of the spiral and its arc-length correspond to
setting $\mu=-1$ and $\lambda=0$ in equation \eqref{eq:integrability_cylinder_solution}. Correspondingly, we have $\tau$ as in \eqref{eq:tau_system_cylinder_solution} in the form
\begin{equation}\label{eq:spiralTau}
\tau=\frac{s}{z-c_0}.
\end{equation}

We chose as the anchor point $\x_0$ on the spiral cylinder $\spicyl$ the point corresponding to $s=1$ and $z=0$ so that $\x_0=\frac{1}{\sqrt{2}}\e_x$ and
the principal frame $\framen$ is
\begin{equation}
	\label{eq:spiral_frame}
	\begin{cases}
		\e_1=\e_z,\\
		\e_2=\tangent(1)=\frac{1}{\sqrt{2}}\left(\e_x+\e_y\right)\\
		\normal=\e_z\times\tangent(1)=\frac{1}{\sqrt{2}}\left(\e_y-\e_x\right).
	\end{cases}
\end{equation} 
Letting $\vt_0$ be the value of $\vt$ at $\x_0$, we see from \eqref{eq:spiralTau} that 
\begin{equation}
	\label{eq:spiral_c_0}
	c_0=-\frac{1}{\tan\vt_0}.
	\end{equation}

With $\kappa$ and $\tau$ as in \eqref{eq:spiralCurvature} and \eqref{eq:spiralTau}, the components of the $\av$ vectors in \eqref{eq:alphas}
are
\begin{align}
 \alpha_1&=\frac{\kappa\tau}{1+\tau^2}=\frac{-(z-c_0)}{s^2+(z-c_0)^2}\,,\\
 \alpha_2&=\frac{\kappa\tau^2}{1+\tau^2}=\frac{-s}{s^2+(z-c_0)^2}\,.
\end{align}

On our surface
\begin{equation}
\rv(s,z)=z\e_z+\rv_0(s)
\end{equation}
we have
\begin{equation}
\dot{\rv}(s,z)=\dot{z}\e_z+\dot{s}\tangent
\end{equation}
for a parameterized curve $\dot{\rv}$ with the dot denoting differentiation with respect to the curve parameter. The non-vanishing entries in the matrix in \eqref{eq:W_matrix} are thus
\begin{equation}\label{eq:spiralW}
	\begin{cases}
\mathbb{W}_{13}= -\av_2\cdot\dot{\rv}=-\alpha_2\dot{z}+\alpha_1\dot{s}=\dfrac{s\dot{z}-(z-c_0)\dot{s}}{s^2+(z-c_0)^2}\,,\\
\mathbb{W}_{23}=-\kappa\tangent\cdot\dot{\rv}+\av_1\cdot\dot{\rv}=\alpha_1\dot{z}+(\alpha_2-\kappa)\dot{s}= \dfrac{-(z-c_0)\dot{z}-s\dot{s}}{s^2+(z-c_0)^2} + \dfrac{\dot{s}}{s}\,.
	\end{cases}
\end{equation}

Since $\cv=\zero$, $\AA$ is path-independent and its non-vanishing entries can easily be found to be
	\begin{equation}
		\label{eq:spiral_A}
		\begin{cases}
			\AA_{13}(s,z)=\arctan\left(\dfrac{z-c_0}{s}\right)-\arctan\left(\dfrac{z_0-c_0}{s_0}\right),\\
			\AA_{23}(s,z)=\dfrac12\ln\dfrac{s^2(s_0^2+(z_0-c_0)^2)}{s_0^2(s^2+(z-c_0)^2)},
		\end{cases}
	\end{equation}
with the prescription that $\AA(s_0,z_0)=0$. The commutation condition \eqref{eq:commutation_condition} reads again as in \eqref{eq:commutation_condition_cylinder} and can be given the form
\begin{equation}
	\label{eq:commutation_condition_spiral}
	[\AA_{13}(z-c_0)+\AA_{23}s][(z-c_0)\dot{s}-s\dot{z}]=0,
\end{equation}
which is satisfied only along the paths described by
\begin{equation}
	\label{eq:spiral_paths}
	z-c_0=s\frac{z_0-c_0}{s_0},
\end{equation}
upon which, incidentally, $\tau$ in \eqref{eq:spiralTau} is constant. 

\begin{figure}[] 
	\centering
	\begin{subfigure}[c]{4.4cm}
		\centering
		\includegraphics[width=4.4cm]{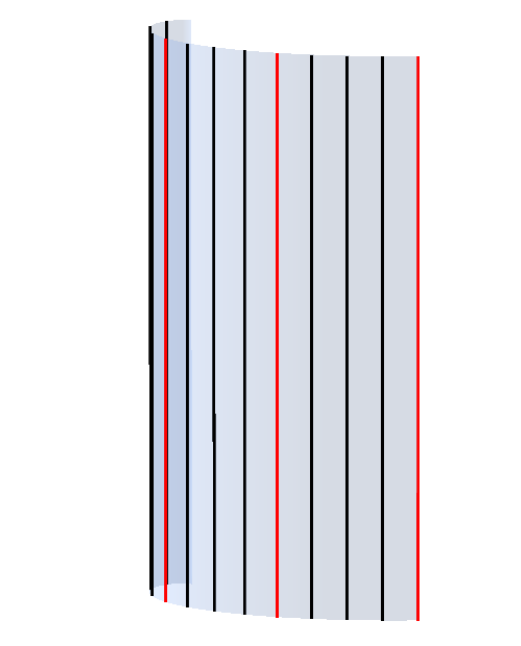}
		\caption{$\vt_0=0$}
		\label{fig:S_theta_0}
	\end{subfigure}
	\qquad
	\begin{subfigure}[c]{4.4cm}
		\centering
		\includegraphics[width=4.4cm]{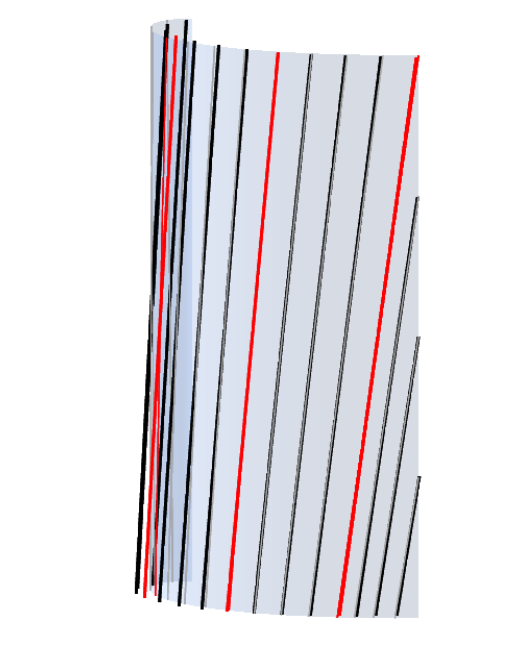}
		\caption{$\vt_0=\frac{\pi}{60}$}
		\label{fig:S_theta_Pio60}
	\end{subfigure}
	\qquad
	\begin{subfigure}[c]{4.4cm}
		\centering
		\includegraphics[width=4.4cm]{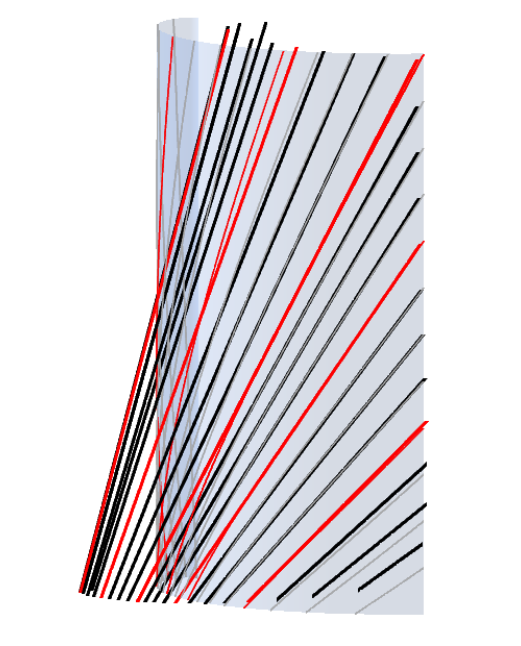}
		\caption{$\vt_0=\frac{\pi}{12}$}
		\label{fig:S_theta_Pio12}
	\end{subfigure}
	\\
	\begin{subfigure}[c]{4.4cm}
		\centering
		\includegraphics[width=4.4cm]{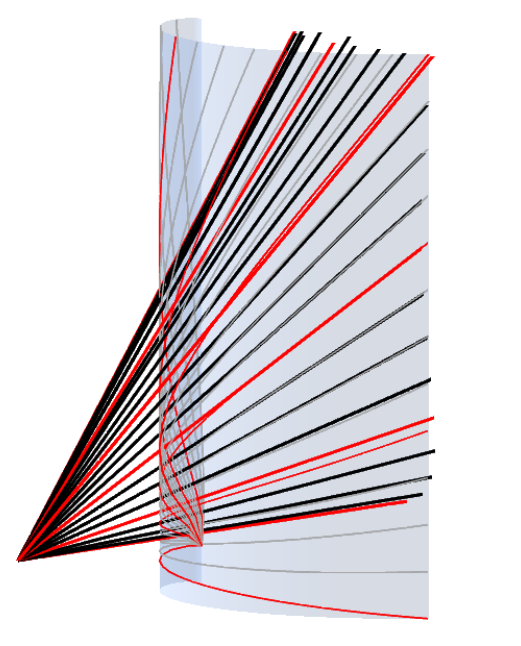}
		\caption{$\vt_0=\frac{\pi}{6}$}
		\label{fig:S_theta_Pio6}
	\end{subfigure}
	\qquad
	\begin{subfigure}[c]{4.4cm}
		\centering
		\includegraphics[width=4.4cm]{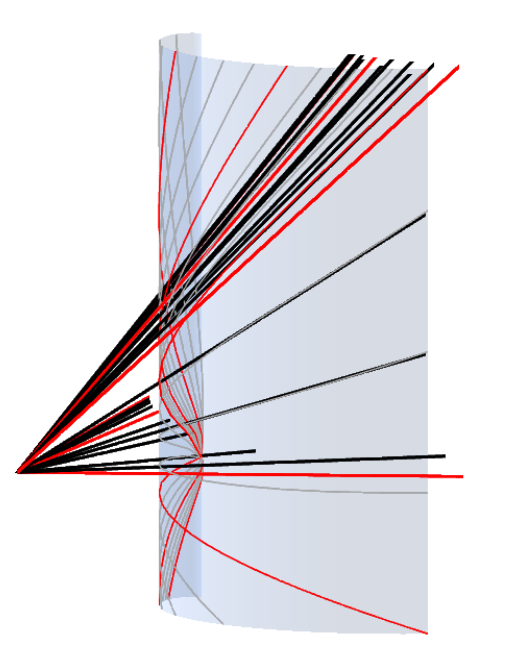}
		\caption{$\vt_0=\frac{\pi}{4}$}
		\label{fig:S_theta_Pio4}
	\end{subfigure}
	\qquad
	\begin{subfigure}[c]{4.4cm}
		\centering
		\includegraphics[width=4.4cm]{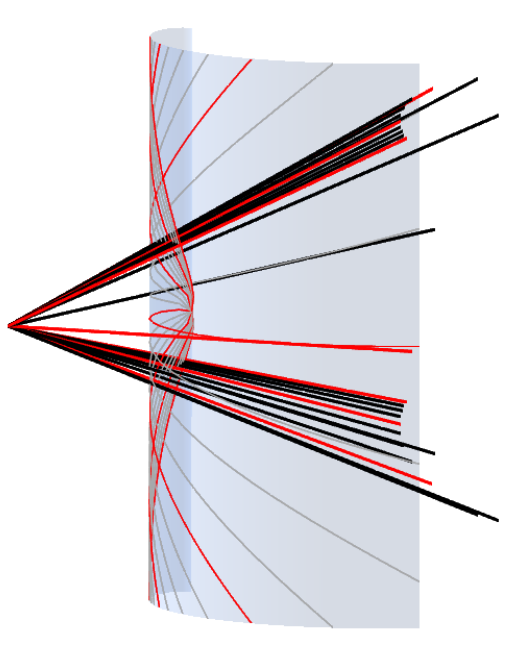}
		\caption{$\vt_0=\frac{\pi}{2}$}
		\label{fig:S_theta_Pio2}
	\end{subfigure}
	\caption{Configurations $\spicyl^\ast(\vt_0)$ traversed by the soft mode of a spiral cylinder $\spicyl$, depicted in the background. To aid the eye, every fourth line is coloured in red. The heavier red and black lines are straight lines, obtained from \eqref{eq:spiral_ruled}, illustrating the ruled surfaces $\spicyl^\ast(\vt_0)$. The lighter red and dark gray lines are the preimages on $\spicyl$ of those same lines. The surface $\spicyl^\ast(\vt_0)$ is wrapped around a spiral cone  whose vertex, according to \eqref{eq:cone_vertex} and \eqref{eq:spiral_c_0}, has $z$-coordinate equal to  $-1/\tan\vt_0$.}
	\label{fig:cone}
\end{figure}

Thus, the exponential representation of the moving frames $\framev$ in \eqref{eq:v_solution} is valid only along a single path issued from any given point. However, one can integrate  \eqref{eq:v_differential} on a selected \emph{backbone} path $\x_\mathrm{b}(\zeta):=\x(s_0(\zeta),z_0(\zeta))$ starting from our chosen anchor point $\x_0$ and then apply \eqref{eq:v_solution} to a path joining  any desired point $\x(s,z)$ to the corresponding point on the backbone obtained by inverting \eqref{eq:spiral_paths}. For example, one can choose $s_0(\zeta)\equiv1$ as backbone, derive $z_0(\zeta)$ from \eqref{eq:spiral_paths},
\begin{equation}
	\label{eq:backbone_z_0}
	z_0(\zeta)=c_0+\frac{z-c_0}{s},
\end{equation}
find the frame $\framev$ at $\x(1,z_0(\zeta))$ by solving  \eqref{eq:v_differential}, and then finally arrive at
\begin{equation}
	\label{eq:spiral_frame_final}
	\begin{pmatrix}
		\vv_1\\
		\vv_2\\
		\normal^\ast
	\end{pmatrix}(\x(s,z))=
	\ex^{\AA(s,z)}
	\left.\begin{pmatrix}
		\vv_1\\
		\vv_2\\
		\normal^\ast
	\end{pmatrix}\right|_{\x(1,z_0(\zeta))},
\end{equation}
where $\AA(s,z)$ is as in \eqref{eq:spiral_A} with $z_0=z_0(\zeta)$. A similar strategy can be applied by choosing as backbone the path $z_0(\zeta)\equiv0$.
\begin{remark}
	\label{rmk:ruled_surface}
Actually,  in the specific case under study, there is a far easier way to represent the frame $\framev$ at $\x(s,z)$, a way with a transparent geometric meaning. Let $\rv_\mathrm{b}$ denote the path \eqref{eq:spiral_paths} on $\spicyl$ passing through the backbone point $\x_\mathrm{b}(\zeta)$. By \eqref{eq:commutation_condition_spiral} and \eqref{eq:spiral_paths}, we can write $\dot{\rv}_\mathrm{b}=\dot{z}\e_z+\dot{s}\tangent$ with
\begin{equation}
\label{eq:spiral_path_dot}
\dot{z}=\frac{z_0-c_0}{s_0}\dot{s}.
\end{equation}
Use of \eqref{eq:spiral_path_dot} and \eqref{eq:spiral_paths} in \eqref{eq:spiralW} shows that $\WW(\dot{\rv}_\mathrm{b})=0$, and so by \eqref{eq:v_differential} the frame $\framev$ is constant along $\rv_\mathrm{b}$. It readily follows from the integration along $\rv_\mathrm{b}$ in \eqref{eq:y_solution} that
\begin{equation}
	\label{eq:spiral_ruled}
	\begin{split}
	\y(s,z)&=\y(\x_\mathrm{b})+\left[\frac{z_0-c_0}{s_0}\vv_1(\x_\mathrm{b})+\vv_2(\x_\mathrm{b})\right](s-s_0)\\
	&=\y(\x_\mathrm{b})+\left[\vv_1(\x_\mathrm{b})+\frac{s_0}{z_0-c_0}\vv_2(\x_\mathrm{b})\right](z-z_0),
	\end{split}
\end{equation}
where the curve $(s_0(\zeta),z_0(\zeta))$ is expressed in terms of $(s,z)$ by inverting  \eqref{eq:spiral_paths} with respect to $\zeta$, as done  in \eqref{eq:backbone_z_0} for $s_0(\zeta)\equiv1$.
\end{remark}
\begin{remark}
\label{rmk:cone_rev}
Irrespective of the point on the backbone $\x_\mathrm{b}$, all lines in \eqref{eq:spiral_paths} pass through the point with $s=0$ and $z=c_0$; they are spirals on $\spicyl$ that are mapped into the straight lines in space represented by \eqref{eq:spiral_ruled}. These go all through one and the same vertex, which is the image of the point with $s=0$ and $z=c_0$. Thus, the surface $\spicyl^\ast$ is wrapped around a generalized cone with vertex identified by the vector
	\begin{equation}
	\label{eq:cone_vertex}
	\av_0=\x_0+c_0\e_z-\tangent(1),
\end{equation}
where $\x_0$ is the anchor point, as readily follows from \eqref{eq:spiral_ruled} by setting  $s_0=1$, $z_0=0$, $s=0$, and $z=c_0$.
\end{remark}
\begin{remark}
	\label{rmk:roussos_rev}
	Let $\Gamma$ be the image on $\spicyl$ of the straight segment joining in the $(s,z)$ plane the points $(0,c_0)$ and $(s,z)$. Since the deformation in \eqref{eq:spiral_ruled} is an isometry, it maps  $\Gamma$  into a straight segment in space of equal length,
	\begin{subequations}\label{eq:roussos}
	\begin{equation}
		\label{eq:roussos_lambda}
		\lambda(s,z)=\sqrt{(z-c_0)^2+s^2}.
	\end{equation}
Let $\Gamma_\perp$ be the curve orthogonal on $\spicyl$ to $\Gamma$  that corresponds to the arc of unit circle in the $(s,z)$ plane spanned by the half-line $s=0$, $z\geqq c_0$  and the  preimage  of $\Gamma$ in that plane. The isometry maps $\Gamma_\perp$ into a curve on the unit sphere of length
\begin{equation}
	\label{eq:roussos_ell}
	\ell(s,z)=\arcsin\frac{s}{\sqrt{(z-c_0)^2+s^2}}.
\end{equation}
\end{subequations}
Equations \eqref{eq:roussos} describe precisely the one-parameter family of isometries envisaged by \citet{roussos:principal} between a spiral cylinder $\spicyl$ and a spiral cone $\cone$.\footnote{See, in particular, equation (8) of \citep{roussos:principal}, where a typo should be corrected to bring it to agree with \eqref{eq:roussos_ell}.} Thus, the soft mode of a spiral cylinder indeed wraps it around a spiral cone.
\end{remark}
Figure~\ref{fig:cone} illustrates  for several values of $\vt_0$ in \eqref{eq:spiral_c_0} how $\spicyl^\ast(\vt_0)$ is wrapped around a spiral cone $\cone$.
Shown in every case is the patch
of $\spicyl$ corresponding to the parameter range $(s,z)\in(0,3/2]\times[-2,2]$ along with
the corresponding deformed surface. In panels (d)--(f) of  Fig.~\ref{fig:cone}, where $c_0=-1/\tan\vt_0\in[-2,2]$, the vertex of $\cone$, whose position is given by \eqref{eq:cone_vertex}, falls inside the image of the patch, whereas in the other panels it falls outside. In the limit as $c_0\to-\infty$, the vertex of $\cone$ becomes the point at infinity and $\spicyl^\ast$ reduces to $\spicyl$.\footnote{By symmetry, in the limit as $c_0\to+\infty$ the vertex of $\cone$ escapes again to infinity, but on the opposite side, while $\spicyl^\ast$ still reduces to $\spicyl$.}

\begin{figure}[] 
\centering
\begin{subfigure}[b]{6.4cm}
	\centering
\includegraphics[width=\linewidth]{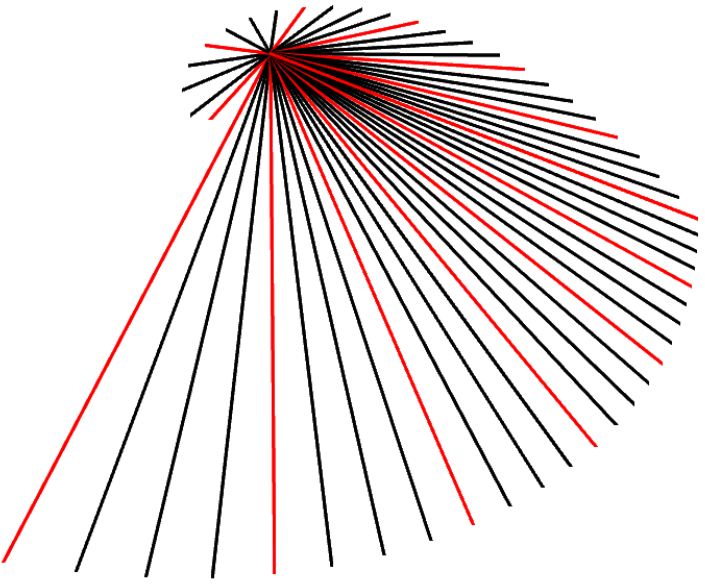}
\caption{Top view}
\end{subfigure}
\qquad\qquad
\begin{subfigure}[b]{4.8cm}
	\centering
	\includegraphics[width=\linewidth]{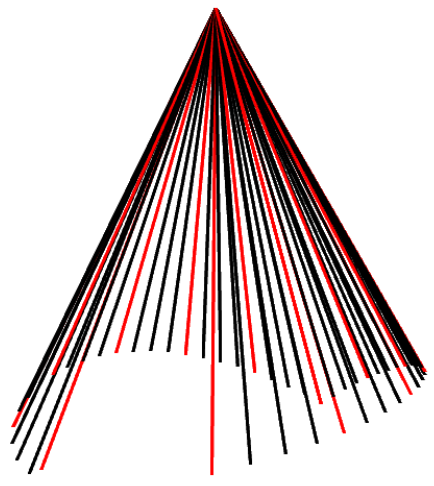}
	\caption{Side view}
	\end{subfigure}
\caption{Two views of the surface $\spicyl^\ast(\frac{\pi}{6})$.}
\label{fig:soleSpiral}
\end{figure}

In Fig.~\ref{fig:soleSpiral} two views of $\spicyl^\ast(\frac{\pi}{6})$ show how the straight lines emanating from the vertex of $\cone$ wind around it reflecting the spiral nature they share with the surface of origin.

We have seen that the soft mode of a single spiral cylinder comprises a family of spiral cones. Conversely, the soft mode of a spiral cone comprises other spiral cones and a limiting spiral cylinder.

\section{Closing comments}\label{sec:conclusion}
Soft modes of a shell are (one-parameter) families of deformations of a reference configuration that cost no elastic energy. One could envision the configurations reached by a soft mode as populating a rather crowded ground state alongside their reference configuration. Every elastic theory  has its own soft modes and reference configurations admitting soft modes could be different in different theories.

In \citep{sonnet:variational}, we formulated a variational theory for soft shells characterized by splitting the stored elastic energy density in three independent components, namely, stretching, drilling, and bending. A deformation mode where one single energy component vanishes is said to be neutral. Thus, a soft mode is at the same time stretching, drilling, \emph{and} bending neutral.

Individual neutral modes can interfere with one another. 
In a paper recently published by \citet{tiwari:orthogonal}, the drilling energy  was proved to be fully determined by the metric properties of the deformation. This, in particular, entails that, as also stated by \citet{sonnet:neutral} and proved by \citet{sonnet:metric}, an isometric deformation, which is stretching neutral, is also drilling neutral.

Neutral modes of minimal surfaces have also been studied by \citet{sonnet:neutral}. \citet{tiwari:orthogonal} also showed by example that drilling neutral modes of minimal surfaces are not necessarily bending neutral.\footnote{They pointed out that this implication remains valid for \emph{conformal} deformations of minimal surfaces, which they claim is \emph{implicitly} the case in the setting of \citet{sonnet:neutral}. It should, however, been noted that the deformations considered by \citet{sonnet:neutral} are actually \emph{explicitly} conformal, as also pointed out by \citet{sonnet:addendum}.}

We considered the soft modes predicted by the theory of \citet{sonnet:variational}. We showed that finding all such modes is a problem first posited by Bonnet within the classical theory of surfaces in three-dimensional space. Not all smooth surfaces admit soft modes; those that do form Bonnet families. There are also Bonnet pairs, isolated surfaces that have the same elastic energy, but that are separated by an energy barrier.

While differential geometry deems equivalent all surfaces in one and the same Bonnet family, continuum mechanics is interested in the soft modes of an elastic surface, as they embody the free actuation modes of that surface. 

We gave a criterion to identify soft modes (as well as Bonnet families) and we applied it to a number of exemplary cases that include circular cylinders, spiral cylinders, and spiral cones. Other cases, however, remain to be investigated in the rich realm of Bonnet families. 

We were scarcely concerned with Bonnet pairs. Although these are not associated with soft modes and free fluctuations of shells, they may also be of interest in continuum mechanics, whenever a multi-well theory of shells is needed. Were the individual energy wells provided with soft modes of their own, they would remain separated by an energy barrier.

\appendix

\section{Alternative proof of \citeauthor{tiwari:orthogonal}'s (\citeyear{tiwari:orthogonal}) formula}\label{sec:alternative}
In this appendix, we give a proof phrased in the formalism adopted in this paper of equation \eqref{eq:a_3_fried} in the main text due to \citet{tiwari:orthogonal}.

By use of \eqref{eq:structure_laws}, a tedious but easy calculation that starts from differentiating both sides of \eqref{eq:U} readily leads us to
\begin{equation}
\begin{split}
\label{eq:a_3_fried_step_1}
\U\nablas\U&=\lambda_1\uv_1\otimes\uv_1\otimes\nablas\lambda_1+\lambda_1\lambda_2\uv_2\otimes\uv_1\otimes\cv+\lambda_1^2\uv_1\otimes\uv_2\otimes\cv+\lambda_1^2\uv_1\otimes\normal\otimes\dv_1\\
&+\lambda_2\uv_2\otimes\uv_2\otimes\nablas\lambda_2-\lambda_1\lambda_2\uv_1\otimes\uv_2\otimes\cv-\lambda_2^2\uv_2\otimes\uv_1\otimes\cv+\lambda_2^2\uv_2\otimes\normal\otimes\dv_2.
\end{split}
\end{equation}
By the right circle contraction defined in \eqref{eq:circles_definition}, it follows from \eqref{eq:a_3_fried_step_1} that
\begin{equation}
	\label{eq:a_3_fried_step_2}
	\U\nablas\U\circ\W(\normal)=-\lambda_1[(\lambda_2-\lambda_1)(\cv\cdot\uv_1)+\nablas\lambda_1\cdot\uv_2]\uv_1-\lambda_2[(\lambda_1-\lambda_2)(\cv\cdot\uv_2)-\nablas\lambda_2\cdot\uv_1]\uv_2,
\end{equation}
which by \eqref{eq:integrability_condition_2} and \eqref{eq:integrability_condition_3} implies \eqref{eq:a_3_fried}.

\section{Sufficient condition for path-independence}\label{sec:c=0}
As also recalled, among many others, by \citet{sonnet:bending-neutral},\footnote{See also \citep[p.\,243]{weatherburn:differential_1}.} a sufficient and necessary condition for a smooth \emph{tangential} vector field $\vv$ on $\surface$ to be the surface gradient of a scalar field $\psi$, $\vv=\nablas\psi$, is 
\begin{equation}\label{eq:integrability_scalar}
\curls\vv=\normal\times\curvature\vv,
\end{equation}
which is the vectorial form of the integrability condition \eqref{eq:integrability_vector} and, by \eqref{eq:curl_tangential_component}, reads equivalently as
\begin{equation}
	\label{eq:curl_c}
	\curls\vv\cdot\normal=0.
\end{equation}
Our proof that \eqref{eq:c=0} suffices to guarantee that the matrix $\AA(\rv)$ in \eqref{eq:A_matrix} is independent of the integration path $\rv$ will be entirely based on \eqref{eq:curl_c}.

We start from the entry $\WW_{12}$ of $\WW$ in \eqref{eq:W_matrix},
\begin{equation}
	\label{eq:W_12}
	\WW_{12}=\cv\cdot\dot{\rv}.
\end{equation}
The corresponding entry $\AA_{12}$ of $\AA$ is path-independent if and only if $\cv=\nablas\psi_{12}$, so that $\WW_{12}=\dot{\psi}_{12}$. By applying \eqref{eq:curl_c} to $\cv$, by \eqref{eq:gauss_equation}, we easily find the condition
\begin{equation}\label{eq:K=0}
	K=0.
\end{equation}

Similarly, there exists a scalar function $\psi_{13}$ such that 
\begin{equation}
	\label{eq:W_13}
	\WW_{13}=-(\kappa_1\e_1+\av_2)\cdot\dot{\rv}=\dot{\psi}_{13},
\end{equation}
if and only if
\begin{equation}
	\label{eq:curl_a_2}
	\tfrac12c_1(\kappa_2+\kappa_1)+\tfrac12c_2(\kappa_2-\kappa_1)\cos2\vt+\tfrac12c_2(\kappa_2-\kappa_1)\sin2\vt=0,
\end{equation}
where we have set $\cv=c_1\e_1+c_2\e_2$ and use has been made of \eqref{eq:structure_laws_curvature}, \eqref{eq:integrability_conditions_rotation_bis_3}, \eqref{eq:alphas}, and \eqref{eq:alphas_representation}.

In precisely the same way, we proceed by requiring the existence of a scalar field $\psi_{23}$ such that 
\begin{equation}
	\label{eq:W_23}
	\WW_{23}=-(\kappa_2\e_2-\av_1)\cdot\dot{\rv}=\dot{\psi}_{23}. 
\end{equation}
By using \eqref{eq:integrability_conditions_rotation_bis_3} instead of \eqref{eq:integrability_conditions_rotation_bis_2} and using all other equations that led us to \eqref{eq:curl_a_2}, we arrive at
\begin{equation}
	\label{eq:curl_a_1}
		\tfrac12c_2(\kappa_2+\kappa_1)+\tfrac12c_1(\kappa_2-\kappa_1)\sin2\vt-\tfrac12c_2(\kappa_2-\kappa_1)\cos2\vt=0.
\end{equation}

It is thus clear that $\cv=\zero$ implies that equations \eqref{eq:curl_c}, \eqref{eq:curl_a_2} and \eqref{eq:curl_a_1} are satisfied, and so $\AA$ is path-independent. However, this is not in general a necessary condition, as the linear system \eqref{eq:curl_a_2}, \eqref{eq:curl_a_1} in the variables $(c_1,c_2)$ has determinant $K$, and so it may also have non-vanishing solutions when \eqref{eq:K=0} applies. On the other hand, \eqref{eq:K=0} is always a necessary condition for $\AA$ to be path-independent.


\begin{thebibliography}{38}
	\expandafter\ifx\csname natexlab\endcsname\relax\def\natexlab#1{#1}\fi
	\providecommand{\url}[1]{\texttt{#1}}
	\providecommand{\href}[2]{#2}
	\providecommand{\path}[1]{#1}
	\providecommand{\DOIprefix}{doi:}
	\providecommand{\ArXivprefix}{arXiv:}
	\providecommand{\URLprefix}{URL: }
	\providecommand{\Pubmedprefix}{pmid:}
	\providecommand{\doi}[1]{\href{http://dx.doi.org/#1}{\path{#1}}}
	\providecommand{\Pubmed}[1]{\href{pmid:#1}{\path{#1}}}
	\providecommand{\bibinfo}[2]{#2}
	\ifx\xfnm\relax \def\xfnm[#1]{\unskip,\space#1}\fi
	\bibitem[{Bobenko and Eitner(1998)}]{bobenko:bonnet}
	\bibinfo{author}{Bobenko, A.}, \bibinfo{author}{Eitner, U.},
	\bibinfo{year}{1998}.
	\newblock \bibinfo{title}{Bonnet surfaces and {P}ainlev\'e equations}.
	\newblock \bibinfo{journal}{J. reine angew. Math.} \bibinfo{volume}{499},
	\bibinfo{pages}{47--81}.
	\newblock \DOIprefix\doi{doi:10.1515/crll.1998.061}.
	\bibitem[{Bobenko and Eitner(2000)}]{bobenko:painleve}
	\bibinfo{author}{Bobenko, A.I.}, \bibinfo{author}{Eitner, U.},
	\bibinfo{year}{2000}.
	\newblock \bibinfo{title}{Painlev\'e Equations in the Differential Geomerty of
		Surfaces}. volume \bibinfo{volume}{1753} of \textit{\bibinfo{series}{Lecture
			Notes in Mathematics}}.
	\newblock \bibinfo{publisher}{Springer}, \bibinfo{address}{Berlin}.
	\bibitem[{Bobenko et~al.(2025)Bobenko, Hoffmann and
		Sageman-Furnas}]{bobenko:compact}
	\bibinfo{author}{Bobenko, A.I.}, \bibinfo{author}{Hoffmann, T.},
	\bibinfo{author}{Sageman-Furnas, A.O.}, \bibinfo{year}{2025}.
	\newblock \bibinfo{title}{Compact bonnet pairs: isometric tori with the same
		curvatures}.
	\newblock \bibinfo{journal}{Publ. Math. {IHES}} \bibinfo{volume}{142},
	\bibinfo{pages}{241--293}.
	\newblock \DOIprefix\doi{https://doi.org/10.1007/s10240-025-00159-z}.
	\bibitem[{Bonnet(1867)}]{bonnet:memoire_II}
	\bibinfo{author}{Bonnet, O.}, \bibinfo{year}{1867}.
	\newblock \bibinfo{title}{M{\'e}moire sur la th{\'e}orie des surfaces
		applicables sur une surface donn{\'e}e. {II.} {D}{\'e}termination de toutes
		les surfaces applicables sur une surface donn{\'e}e.}
	\newblock \bibinfo{journal}{J. {\'E}cole Pol. Paris} \bibinfo{volume}{25},
	\bibinfo{pages}{1--151}.
	\newblock \bibinfo{note}{Available from
		\url{https://gallica.bnf.fr/ark:/12148/bpt6k433698b/f5}}.
	\bibitem[{Cartan(1942)}]{cartan:couples}
	\bibinfo{author}{Cartan, {\'E}.}, \bibinfo{year}{1942}.
	\newblock \bibinfo{title}{Sur les couples de surface applicables avec
		conservation des courbures principales}.
	\newblock \bibinfo{journal}{Bull. Sc. Math.} \bibinfo{volume}{66},
	\bibinfo{pages}{55--85}.
	\newblock \bibinfo{note}{Available from
		\url{https://gallica.bnf.fr/ark:/12148/bpt6k9632895z/f61}}.
	\bibitem[{Chern(1985)}]{chern:deformation}
	\bibinfo{author}{Chern, S.S.}, \bibinfo{year}{1985}.
	\newblock \bibinfo{title}{Deformation of surfaces preserving principal
		curvatures}, in: \bibinfo{editor}{Chavel, I.}, \bibinfo{editor}{Farkas, H.M.}
	(Eds.), \bibinfo{booktitle}{Differential Geometry and Complex Analysis. {A}
		volume dedicated to the memory of {H}arry {E}rnest {R}auch}.
	\bibinfo{publisher}{Springer-Verlag}, \bibinfo{address}{Berlin}, pp.
	\bibinfo{pages}{155--164}.
	\bibitem[{Colares and Kenmotsu(1989)}]{colares:isometric}
	\bibinfo{author}{Colares, A.G.}, \bibinfo{author}{Kenmotsu, K.},
	\bibinfo{year}{1989}.
	\newblock \bibinfo{title}{Isometric deformation of surfaces in ${R}^3$
		preserving the mean curvature function}.
	\newblock \bibinfo{journal}{Pacific J. Math.} \bibinfo{volume}{136},
	\bibinfo{pages}{71--80}.
	\newblock \DOIprefix\doi{10.2140/pjm.1989.136.71}.
	\bibitem[{Cutts(2026)}]{cutts:twisty}
	\bibinfo{author}{Cutts, E.}, \bibinfo{year}{2026}.
	\newblock \bibinfo{title}{Two twisty shapes resolve a centuries-old topology
		puzzle}.
	\newblock \bibinfo{note}{Available from
		\url{https://www.quantamagazine.org/two-twisty-shapes-resolve-a-centuries-old-topology-puzzle-20260120/}}.
	\bibitem[{Dierkes et~al.(2010)Dierkes, Hildebrandt and
		Sauvigny}]{dierkes:minimal}
	\bibinfo{author}{Dierkes, U.}, \bibinfo{author}{Hildebrandt, S.},
	\bibinfo{author}{Sauvigny, F.}, \bibinfo{year}{2010}.
	\newblock \bibinfo{title}{Minimal Surfaces}. volume \bibinfo{volume}{339} of
	\textit{\bibinfo{series}{A Series of Comprehensive Studies in Mathematics}}.
	\newblock \bibinfo{edition}{2nd} ed., \bibinfo{publisher}{Springer},
	\bibinfo{address}{Berlin}.
	\bibitem[{Golubovi\'c and Lubensky(1989)}]{golubovic:nonlinear}
	\bibinfo{author}{Golubovi\'c, L.}, \bibinfo{author}{Lubensky, T.C.},
	\bibinfo{year}{1989}.
	\newblock \bibinfo{title}{Nonlinear elasticity of amorphous solids}.
	\newblock \bibinfo{journal}{Phys. Rev. Lett.} \bibinfo{volume}{63},
	\bibinfo{pages}{1082--1085}.
	\bibitem[{Hughes and Brezzi(1989)}]{hughes:drilling}
	\bibinfo{author}{Hughes, T.J.R.}, \bibinfo{author}{Brezzi, F.},
	\bibinfo{year}{1989}.
	\newblock \bibinfo{title}{On drilling degrees of freedom}.
	\newblock \bibinfo{journal}{Comput. Methods Appl. Mech. Engrg.}
	\bibinfo{volume}{72}, \bibinfo{pages}{105--121}.
	\newblock \DOIprefix\doi{10.1016/0045-7825(89)90124-2}.
	\bibitem[{Hurewicz(2014)}]{hurewicz:lectures}
	\bibinfo{author}{Hurewicz, W.}, \bibinfo{year}{2014}.
	\newblock \bibinfo{title}{Lectures on Ordinary Differential Equations}.
	\newblock \bibinfo{publisher}{Dover}, \bibinfo{address}{Mineola}.
	\bibitem[{Jensen et~al.(2016)Jensen, Musso and Nicolodi}]{jensen:surfaces}
	\bibinfo{author}{Jensen, G.R.}, \bibinfo{author}{Musso, E.},
	\bibinfo{author}{Nicolodi, L.}, \bibinfo{year}{2016}.
	\newblock \bibinfo{title}{Surfaces in Classical Geometries. {A} Treatment by
		Moving Frames}.
	\newblock Universitext, \bibinfo{publisher}{Springer}, \bibinfo{address}{Cham}.
	\bibitem[{Jensen et~al.(2018)Jensen, Musso and Nicolodi}]{jensen:compact}
	\bibinfo{author}{Jensen, G.R.}, \bibinfo{author}{Musso, E.},
	\bibinfo{author}{Nicolodi, L.}, \bibinfo{year}{2018}.
	\newblock \bibinfo{title}{Compact surfaces with no {B}onnet mate}.
	\newblock \bibinfo{journal}{J. Geom. Anal.} \bibinfo{volume}{28},
	\bibinfo{pages}{2644--2652}.
	\newblock \DOIprefix\doi{https://doi.org/10.1007/s12220-017-9924-y}.
	\bibitem[{Lawson and Tribuzy(1981)}]{lawson:mean}
	\bibinfo{author}{Lawson, H.B.}, \bibinfo{author}{Tribuzy, R.},
	\bibinfo{year}{1981}.
	\newblock \bibinfo{title}{On the mean curvature function for compact surfaces}.
	\newblock \bibinfo{journal}{J. Differ. Geom.} \bibinfo{volume}{24},
	\bibinfo{pages}{179--183}.
	\newblock \DOIprefix\doi{10.4310/jdg/1214436095}.
	\bibitem[{Man and Cohen(1986)}]{man:coordinate}
	\bibinfo{author}{Man, C.S.}, \bibinfo{author}{Cohen, H.}, \bibinfo{year}{1986}.
	\newblock \bibinfo{title}{A coordinate-free approach to the kinematics of
		membranes}.
	\newblock \bibinfo{journal}{J. Elast.} \bibinfo{volume}{16},
	\bibinfo{pages}{97--104}.
	\newblock \DOIprefix\doi{10.1007/BF00041068}.
	\bibitem[{Marsden and Ratiu(1999)}]{marsden:introduction}
	\bibinfo{author}{Marsden, J.E.}, \bibinfo{author}{Ratiu, T.S.},
	\bibinfo{year}{1999}.
	\newblock \bibinfo{title}{Introduction to Mechanics and Symmetry. {A} Basic
		Exposition of Classical Mechanical Systems}. volume~\bibinfo{volume}{17} of
	\textit{\bibinfo{series}{Texts in Applied Mathematics}}.
	\newblock \bibinfo{edition}{2nd} ed., \bibinfo{publisher}{Springer},
	\bibinfo{address}{New York}.
	\bibitem[{Needham(2021)}]{needham:visual}
	\bibinfo{author}{Needham, T.}, \bibinfo{year}{2021}.
	\newblock \bibinfo{title}{Visual Differential Geometry and Forms. {A}
		mathematical drama in five acts}.
	\newblock \bibinfo{publisher}{Princeton University Press},
	\bibinfo{address}{Princeton}.
	\bibitem[{Nirenberg(2018)}]{nirenberg:lectures}
	\bibinfo{author}{Nirenberg, L.}, \bibinfo{year}{2018}.
	\newblock \bibinfo{title}{Lectures on Differential Equations and Differential
		Geometry}. volume~\bibinfo{volume}{7} of \textit{\bibinfo{series}{Classical
			Topics in Mathematics}}.
	\newblock \bibinfo{publisher}{Higher Education Press},
	\bibinfo{address}{Beijing}.
	\bibitem[{Olmsted(1994)}]{olmsted:rotational}
	\bibinfo{author}{Olmsted, P.D.}, \bibinfo{year}{1994}.
	\newblock \bibinfo{title}{Rotational invariance and {G}oldstone modes in
		nematic elastomers and gels}.
	\newblock \bibinfo{journal}{J. Phys. II France} \bibinfo{volume}{4},
	\bibinfo{pages}{2215--2230}.
	\bibitem[{Pressley(2012)}]{pressley:elementary}
	\bibinfo{author}{Pressley, A.}, \bibinfo{year}{2012}.
	\newblock \bibinfo{title}{Elementary Differential Geometry}.
	\newblock Springer Undergraduate Mathematics Series. \bibinfo{edition}{2nd}
	ed., \bibinfo{publisher}{Springer-Verlag}, \bibinfo{address}{London}.
	\bibitem[{Rodrigues(1840)}]{rodrigues:lois}
	\bibinfo{author}{Rodrigues, O.}, \bibinfo{year}{1840}.
	\newblock \bibinfo{title}{Des lois g{\'e}om{\'e}triques qui r{\'e}gissent les
		d{\'e}placements d'un syst{\`e}me solide dans l'espace, et de la variation
		des coordonn{\'e}es provenant de ces d{\'e}placements consid{\'e}r{\'e}s
		ind{\'e}pendamment des causes qui peuvent les produire}.
	\newblock \bibinfo{journal}{J. Math. Pures Appl.} \bibinfo{volume}{5},
	\bibinfo{pages}{380--440}.
	\newblock \bibinfo{note}{Available from
		\url{http://www.numdam.org/item/JMPA_1840_1_5__380_0/}}.
	\bibitem[{Roussos(1987)}]{roussos:principal}
	\bibinfo{author}{Roussos, I.M.}, \bibinfo{year}{1987}.
	\newblock \bibinfo{title}{Principal-curvature-preserving isometries of surfaces
		in ordinary space}.
	\newblock \bibinfo{journal}{Bol. Soc. Bras. Mat.} \bibinfo{volume}{18},
	\bibinfo{pages}{95--105}.
	\newblock \DOIprefix\doi{https://doi.org/10.1007/BF02590026}.
	\bibitem[{Roussos(1988)}]{roussos:helicoidal}
	\bibinfo{author}{Roussos, I.M.}, \bibinfo{year}{1988}.
	\newblock \bibinfo{title}{The helicoidal surfaces as {B}onnet surfaces}.
	\newblock \bibinfo{journal}{Tohoku Math. J.} \bibinfo{volume}{40},
	\bibinfo{pages}{485--490}.
	\newblock \DOIprefix\doi{10.2748/tmj/1178227989}.
	\bibitem[{Roussos(1999a)}]{roussos:global}
	\bibinfo{author}{Roussos, I.M.}, \bibinfo{year}{1999}a.
	\newblock \bibinfo{title}{Global results on {B}onnet surfaces}.
	\newblock \bibinfo{journal}{J Geom.} \bibinfo{volume}{65},
	\bibinfo{pages}{151--168}.
	\newblock \DOIprefix\doi{https://doi.org/10.1007/BF01228686}.
	\bibitem[{Roussos(1999b)}]{roussos:tangential}
	\bibinfo{author}{Roussos, I.M.}, \bibinfo{year}{1999}b.
	\newblock \bibinfo{title}{Tangential developable surfaces as {B}onnet
		surfaces}.
	\newblock \bibinfo{journal}{Acta Math. Sin. (Engl. Ser.)} \bibinfo{volume}{15},
	\bibinfo{pages}{269--276}.
	\newblock \DOIprefix\doi{https://doi.org/10.1007/BF02650670}.
	\bibitem[{Roussos and Hernandez(1990)}]{roussos:number}
	\bibinfo{author}{Roussos, I.M.}, \bibinfo{author}{Hernandez, G.E.},
	\bibinfo{year}{1990}.
	\newblock \bibinfo{title}{On the number of distinct isometric immersions of a
		{R}iemannian surface into ${R}^3$ with given mean curvature}.
	\newblock \bibinfo{journal}{Am. J. Math.} \bibinfo{volume}{112},
	\bibinfo{pages}{71--85}.
	\newblock \bibinfo{note}{Available from
		\url{http://www.jstor.org/stable/2374853}}.
	\bibitem[{Sabitov(2012)}]{sabitov:isometric}
	\bibinfo{author}{Sabitov, I.K.}, \bibinfo{year}{2012}.
	\newblock \bibinfo{title}{Isometric surfaces with a common mean curvature and
		the problem of {B}onnet pairs}.
	\newblock \bibinfo{journal}{Sb. Math.} \bibinfo{volume}{203},
	\bibinfo{pages}{111--152}.
	\newblock \DOIprefix\doi{10.1070/SM2012v203n01ABEH004216}.
	\bibitem[{Sonnet and Virga(2024)}]{sonnet:bending-neutral}
	\bibinfo{author}{Sonnet, A.M.}, \bibinfo{author}{Virga, E.G.},
	\bibinfo{year}{2024}.
	\newblock \bibinfo{title}{Bending-neutral deformations of minimal surfaces}.
	\newblock \bibinfo{journal}{Proc. Roy. Soc. Lond. A} \bibinfo{volume}{480},
	\bibinfo{pages}{20240394}.
	\newblock \DOIprefix\doi{10.1098/rspa.2024.0394}.
	\bibitem[{Sonnet and Virga(2025)}]{sonnet:variational}
	\bibinfo{author}{Sonnet, A.M.}, \bibinfo{author}{Virga, E.G.},
	\bibinfo{year}{2025}.
	\newblock \bibinfo{title}{A variational theory for soft shells}.
	\newblock \bibinfo{journal}{J. Mech. Phys. Solids} \bibinfo{volume}{200},
	\bibinfo{pages}{106132}.
	\newblock \DOIprefix\doi{10.1016/j.jmps.2025.106132}.
	\bibitem[{Sonnet and Virga(2026a)}]{sonnet:addendum}
	\bibinfo{author}{Sonnet, A.M.}, \bibinfo{author}{Virga, E.G.},
	\bibinfo{year}{2026}a.
	\newblock \bibinfo{title}{Addendum to neutral deformation modes of minimal
		surfaces}.
	\newblock \bibinfo{journal}{J. Elast.} \bibinfo{volume}{158},
	\bibinfo{pages}{49}.
	\newblock \DOIprefix\doi{https://doi.org/10.1007/s10659-026-10221-4}.
	\bibinfo{note}{also available from
		\url{https://strathprints.strath.ac.uk/96811/}}.
	\bibitem[{Sonnet and Virga(2026b)}]{sonnet:metric}
	\bibinfo{author}{Sonnet, A.M.}, \bibinfo{author}{Virga, E.G.},
	\bibinfo{year}{2026}b.
	\newblock \bibinfo{title}{Metric implications in the kinematics of surfaces}.
	\newblock \bibinfo{journal}{J. Elast.} \bibinfo{volume}{158},
	\bibinfo{pages}{42}.
	\newblock \DOIprefix\doi{https://doi.org/10.1007/s10659-026-10218-z}.
	\bibitem[{Sonnet and Virga(2026c)}]{sonnet:neutral}
	\bibinfo{author}{Sonnet, A.M.}, \bibinfo{author}{Virga, E.G.},
	\bibinfo{year}{2026}c.
	\newblock \bibinfo{title}{Neutral deformation modes of minimal surfaces}.
	\newblock \bibinfo{journal}{J. Elast.} \bibinfo{volume}{158},
	\bibinfo{pages}{25}.
	\newblock \DOIprefix\doi{https://doi.org/10.1007/s10659-026-10201-8}.
	\bibitem[{Tiwari et~al.(2026)Tiwari, Pandey and Fried}]{tiwari:orthogonal}
	\bibinfo{author}{Tiwari, S.}, \bibinfo{author}{Pandey, A.},
	\bibinfo{author}{Fried, E.}, \bibinfo{year}{2026}.
	\newblock \bibinfo{title}{Orthogonal decomposition of an invariant rotation
		gradient and implications for drilling and bending measures for thin shells}.
	\newblock \bibinfo{journal}{J. Mech. Phys. Solids} \bibinfo{volume}{215},
	\bibinfo{pages}{106731}.
	\newblock \DOIprefix\doi{https://doi.org/10.1016/j.jmps.2026.106731}.
	\bibitem[{Traizet(2020)}]{traizet:construction}
	\bibinfo{author}{Traizet, M.}, \bibinfo{year}{2020}.
	\newblock \bibinfo{title}{Construction of constant mean curvature $n$-noids
		using the {DPW} method}.
	\newblock \bibinfo{journal}{J. reine angew. Math.} \bibinfo{volume}{763},
	\bibinfo{pages}{223--249}.
	\newblock \DOIprefix\doi{doi:10.1515/crelle-2018-0031}.
	\bibitem[{Warner and Terentjev(2003)}]{warner:liquid}
	\bibinfo{author}{Warner, M.}, \bibinfo{author}{Terentjev, E.M.},
	\bibinfo{year}{2003}.
	\newblock \bibinfo{title}{Liquid Crystal Elastomers}. volume
	\bibinfo{volume}{120} of \textit{\bibinfo{series}{International Series of
			Monographs on Physics}}.
	\newblock \bibinfo{publisher}{Oxford University Press}, \bibinfo{address}{New
		York}.
	\bibitem[{Weatherburn(2016)}]{weatherburn:differential_1}
	\bibinfo{author}{Weatherburn, C.E.}, \bibinfo{year}{2016}.
	\newblock \bibinfo{title}{Differential Geometry of Three Dimensions}.
	volume~\bibinfo{volume}{I}.
	\newblock \bibinfo{publisher}{Cambridge University Press},
	\bibinfo{address}{Cambridge}.
	\bibitem[{Wolf(1966)}]{wolf:surfaces}
	\bibinfo{author}{Wolf, J.A.}, \bibinfo{year}{1966}.
	\newblock \bibinfo{title}{Surfaces of constant mean curvature}.
	\newblock \bibinfo{journal}{Proc. Amer. Math. Soc.} \bibinfo{volume}{17},
	\bibinfo{pages}{1103--1111}.
	\newblock \DOIprefix\doi{10.1090/S0002-9939-1966-0203527-3}.
	
\end{thebibliography}
\end{document}